\documentclass{article}

\usepackage{arxiv}

\usepackage[utf8]{inputenc} % allow utf-8 input
\usepackage[T1]{fontenc}    % use 8-bit T1 fonts
\usepackage{hyperref}       % hyperlinks
\usepackage{url}            % simple URL typesetting
\usepackage{booktabs}       % professional-quality tables
\usepackage{amsfonts}       % blackboard math symbols
\usepackage{nicefrac}       % compact symbols for 1/2, etc.
\usepackage{microtype}      % microtypography
\usepackage{lipsum}
\usepackage{graphicx}

\usepackage{amsmath}
\usepackage{setspace}
\usepackage{array}
\usepackage{lineno}
\usepackage{bm}
\usepackage{color}
\usepackage{multirow}
\usepackage{ulem}
\usepackage{amssymb}
\usepackage{lineno}

\usepackage{subfig}

\usepackage{algorithm}%
\usepackage{algorithmicx}%
\usepackage{algpseudocode}%
\usepackage{appendix}
\usepackage{makecell} % 在导言区引入
\usepackage{booktabs}

\title{Unsupervised Posterior Sampling for Seismic Data Recovery via Score-Based Generative Priors}

\author{
 Chuangji Meng \\
  Xi'an Jiaotong University\\
 The National Engineering Research Center for Offshore Oil and Gas Exploration\\
  School of Information and Communications Engineering\\
  Xi'an 710049, China \\
  \texttt{cjmeng@xjtu.edu.cn} \\
   \And
 Jinghuai Gao \\
  Xi'an Jiaotong University\\
  The National Engineering Research Center for Offshore Oil and Gas Exploration\\
  School of Information and Communications Engineering\\
  Xi'an 710049, China \\
  \texttt{jhgao@mail.xjtu.edu.cn} \\
  \And
  Zongben Xu\\
  School of Mathematics and Statistics,\\
  Xi'an Jiaotong University\\
  Xi'an 710049, China.\\
  \texttt{zbxu@mail.xjtu.edu.cn} \\
}

\begin{document}
\maketitle
\begin{abstract}
Seismic data restoration is a fundamental task in seismic exploration, yet remains challenging under complex and unknown degradations. Traditional model-driven or task-specific learning methods often require retraining for each degradation type and fail to generalize effectively to unseen field data. In this work, we introduce an unsupervised Posterior Sampling Framework (PSF) built upon Score-based Generative Models (SGMs) for unified seismic data restoration. PSF leverages the pre-trained unconditional SGMs as a seismic-aware generative prior and derives a generalized conditional score function associated with the forward operator of each inverse problem. This enables posterior sampling across different seismic restoration tasks without retraining or supervision. Additionally, an adaptive noise-level estimation mechanism is incorporated to dynamically regulate the noise suppression strength during sampling, enhancing flexibility under varying signal-to-noise ratios and degradation conditions. Extensive experiments on seismic denoising, interpolation, compressed sensing, and deconvolution demonstrate that PSF delivers high-quality samples and exhibits robust generalization to out-of-distribution data. These results highlight the potential of SGMs as a universal prior for seismic inverse problems and establish PSF as a flexible framework for unsupervised posterior inference across diverse degradation scenarios.
\end{abstract}

% keywords can be removed

\section{Introduction}
Seismic data restoration is a fundamental step in seismic exploration and processing, aiming to recover high-fidelity and structurally informative signals from degraded observations. In practice, seismic records are often degraded by various mechanisms, such as noise contamination, missing traces (subsampling), and convolution with band-limited source wavelets. These factors render the inverse mapping to the underlying clean signals highly ill-posed. Consequently, developing a general and robust framework capable of reconstructing seismic data under diverse degradation scenarios remains a central challenge in geophysical inversion.

Traditional approaches to seismic data restoration can be broadly categorized into model-driven and data-driven methods. Model-driven approaches leverage physical, statistical, or structural priors, such as signal predictability \cite{spitz1991seismic,porsani1999seismic}, low-rank structures \cite{oropeza2011simultaneous,innocent2021robust,chen2016simultaneous}, sparsity \cite{herrmann2000aliased,abma20063d,herrmann2008non,fomel2010seislet,liang2014seismic}, and other regularization-based constraints \cite{fomel2007shaping, chen2020adaptive,zhao2025nonstationary, gao2022incorporating},to formulate inverse problems and guide optimization.
For example, predictive filtering methods assume that seismic signals follow linear or planar events in the spatiotemporal domain and design filters to preserve useful signal components in time or frequency domains . Linear assumptions are also widely applied in seismic interpolation, particularly for regularly sampled surveys. Sparse transform-based methods assume that seismic signals are sparse in certain transform domains, such as wavelet, seislet, or curvelet transforms,  allowing effective separation of signal and noise. Low-rank approaches exploit the low-rank structure of seismic data in transformed or physical domains to recover clean signals . 
Despite their effectiveness and interpretability, these model-driven methods heavily rely on specific prior assumptions, are sensitive to parameter tuning, and are often limited when handling complex noise conditions or unknown degradation processes.

In contrast, data-driven deep learning approaches learn nonlinear seismic priors directly from data, providing new avenues for seismic data restoration and inversion. 
Supervised methods require paired noisy-clean datasets and employ architectures such as encoder-decoder models and cascaded convolutional neural networks, which are typically trained in an end-to-end fashion to directly learn the mapping from corrupted inputs to clean seismic outputs.  Approaches that incorporate additional priors further enhance signal reconstruction by leveraging supplementary information, such as  low-rank components \cite{dong2020new} and  local similarity constraints\cite{geng2022}. Self-supervised and unsupervised methods leverage the statistical properties of data itself, such as Noise2Self \cite{batson2019noise2self} and deep image priors \cite{ulyanov2018deep}, to restore clean signals without labeled data. Generative methods, including those based on generative adversarial networks \cite{wang2020generative} and variational autoencoders \cite{feng2021intelligent}, have demonstrated remarkable performance in recovering fine-scale seismic structures and generating high-fidelity samples.
Nevertheless, they are typically trained for specific tasks and degradation types, making them difficult to generalize to unseen data distributions or new field environments without costly retraining.

Seismic data recovery can be viewed as an inverse problem.
From a Bayesian perspective, recovering the useful signal corresponds to drawing samples from the posterior distribution, which provides a distribution of plausible solutions and naturally facilitates uncertainty quantification.
In this context, posterior sampling offers a principled way to incorporate prior knowledge, while deep generative models provide flexible priors that bridge classical Bayesian inference with modern data-driven approaches. 
Conventional Bayesian seismic inversion relies on sampling techniques, such as Markov chain Monte Carlo, Gibbs sampling, and Langevin-type algorithms, to perform posterior inference and characterize uncertainty \cite{duijndam1988bayesian,sen1996bayesian,izzatullah2020bayesian}.	
These approaches typically operate without leveraging deep generative priors, relying instead on analytic or hand-crafted prior formulations. Recent advances in deep generative modeling introduce data-driven priors that offer more expressive representations of seismic structures, providing an additional pathway for posterior sampling in seismic inversion. 
Early studies employing generative adversarial networks (GAN)\cite{goodfellow2014generative} demonstrated that GAN-based priors enable ultrahigh-resolution or geologically realistic inversion results by constraining the solution space to a learned latent manifold \cite{li2019using,xie2025generative,fang2020deep}. These latent-variable priors integrate naturally with Bayesian posterior sampling, reducing the dimensionality of the model space while preserving structural realism. Variational autoencoders (VAE)\cite{kingma2013auto} have likewise been used to capture geological variability and enable efficient posterior exploration through low-dimensional latent representations\cite{mcaliley2024stochastic}, with recent studies demonstrating that VAE-based dimensionality reduction can further improve uncertainty quantification in nonlinear geophysical inversion \cite{liu2022uncertainty}. Similarly, normalizing-flow-based methods\cite{dinh2014nice} have been adopted for seismic tomography and amplitude-versus-offset inversion\cite{arabpour2025bayesian,zhao2022bayesian}, leveraging invertible architectures to flexibly approximate complex, multimodal posterior distributions with significantly reduced computational cost.  
Collectively, these VAE-, GAN-, and flow-based approaches replace hand-crafted priors with learned generative priors, enabling scalable posterior inference and improved uncertainty quantification in seismic inversion.

More recently, score-based generative models (SGMs) \cite{song2019generative,ho2020denoising} have emerged as a highly flexible and expressive class of generative priors. By directly estimating the gradients of complex, high-dimensional data distributions, SGMs enable principled posterior sampling through stochastic differential equations \cite{song2020score}, providing a unified and broadly applicable framework for Bayesian inversion across diverse seismic inverse problems.
In geophysics, SGMs have demonstrated impressive performance across denoising \cite{peng2024seismic,meng2025posterior}, interpolation \cite{liu2024generative,wei2024seismic,meng2024stochastic,wang2024seisfusion,shi2024generative}, resolution enhancement \cite{zhang2024seisresodiff,yu2025unsupervised}, impedance inversion \cite{chen2025unsupervised}, and seismic imaging \cite{baldassari2024conditional}. Their strength lies in representing highly flexible priors that can be adapted to multiple inverse problems by incorporating forward operators during posterior sampling, without requiring task-specific retraining.
	
Nevertheless, existing SGM-based methods remain limited in flexibility and robustness. Many are trained for specific degradation types or tasks, necessitating retraining when the forward operator changes \cite{liu2024generative,zhang2024seisresodiff,baldassari2024conditional}. Conditional score functions are often designed for specific tasks, and few methods provide systematic strategies for handling unknown or variable noise levels, which is critical for field seismic data applications \cite{liu2024generative,wei2024seismic,chen2025unsupervised,wang2024seisfusion,shi2024generative,baldassari2024conditional}. While pre-training and fine-tuning paradigms can adapt generative models to new tasks, they still require supervised training with paired data or self-supervised learning during fine-tuning, and thus do not constitute fully unsupervised posterior sampling. Moreover, during the sampling stage, these generative models cannot simultaneously perform denoising and inversion\cite{cheng2025generative}.

To address these challenges, we propose an unsupervised Posterior Sampling Framework (PSF) for seismic data restoration. PSF treats a pre-trained unconditional SGM as a universal generative prior and constructs a generalized posterior sampling scheme compatible with arbitrary forward operators. This allows flexible restoration of seismic signals under diverse degradation scenarios without retraining. Moreover, PSF integrates an automatic noise-level estimation mechanism to adaptively control the strength of noise suppression during sampling, ensuring robust performance across datasets with varying signal-to-noise ratios. By unifying unconditional SGMs with seismic-aware forward operators, PSF provides a theoretically grounded, flexible, and generalizable framework for seismic inverse problems.

The main contributions of this work are summarized as follows:

\begin{itemize}
	\item [1)]{
We propose a unified unsupervised posterior sampling framework for seismic data recovery, leveraging pre-trained score-based generative models without task-specific retraining.}

\item [2)]{
We enable flexible and memory-efficient adaptation to diverse inverse problems by leveraging a generalized conditional score function linking the seismic-aware generative prior with arbitrary forward operators.}

\item [3)]{
We introduce an automatic noise-level estimation mechanism to facilitate posterior sampling under unknown signal-to-noise ratios.}

\end{itemize}

The remainder of this paper is organized as follows. First, we review score-based generative models and unconditional sampling for seismic data. Next, we present an unsupervised posterior sampling framework based on SGMs, which bridges the seismic-aware generative prior with arbitrary forward operators. Then, we validate our method through numerical examples. Furthermore, we discuss some potential advantages  and limitations of our method. 
Finally, we share our concluding remarks.

\section{Method}   

\subsection{preliminary}

\subsubsection{Score-based generative model}

SGMs are based on estimating and sampling from the score of the data probability density ${p_{{\rm{data}}}}({\bf{x}})$, which is the gradient of the log-density function at the input data point 
${\nabla _{\bf{x}}}\log {p_{data}}({\bf{x}})$:
\begin{equation}
	{{\bf{s}}}({\bf{x}})={\nabla _{{\bf{x}}}}\log {p_{{\sigma _i}}}({\bf{x}})
	\label{Eq:score def}
\end{equation}
In order to better gradually learn the data distribution, SGMs perturb the data using various levels of noise; and simultaneously estimating scores corresponding to all noise levels . The purpose of this is to create an intermediate distribution with a transition from the prior distribution to the target distribution. Gradually adding noise can avoid the situation where the score cannot be estimated when the data density is close to 0. Let 
${p_\sigma }(\widetilde {\bf{x}}\mid {\bf{x}}) = {\cal N}\left( {\widetilde {\bf{x}}\mid {\bf{x}},{\sigma ^2}{\bf{I}}} \right)$  be a perturbation kernel, and denote the corresponding perturbed data distribution as 
${p_\sigma }(\widetilde {\bf{x}}) \buildrel \Delta \over = \int {{p_\sigma }} (\widetilde {\bf{x}}\mid {\bf{x}}){p_{{\rm{data }}}}({\bf{x}}){\rm{d}}{\bf{x}}$. Consider a sequence of noise scales 
$\left\{ {{\sigma _i}} \right\}_{i = 1}^L$  that satisfies 
${\sigma _{\max }} = {\sigma _1} > {\sigma _2} >  \cdots  > {\sigma _L} = {\sigma _{\min }}$, where the sequence is usually geometric or uniform. Typically, $\sigma _{\min }$ is small enough such that ${p_{{\sigma _{\min }}}}({\bf{x}}) \approx {p_{{\rm{data }}}}({\bf{x}})$    , and  
${\sigma _{\max }}$ is large enough such that  
${p_{{\sigma _{\max }}}}({\bf{x}})\approx {\mathcal N}\left( {{\bf{x}}\mid {\bf{0}},{\sigma _{max}}^2{\bf{I}}} \right)$  .

\subsubsection{Pretrined score function of seismic data}
When SGMs use Langevin dynamics for sampling, it needs to know the scores corresponding to all noise levels.  This score function is approximated by a network called conditional score network (NCSN)  based on the denoising score matching algorithm \cite{song2019generative}.
We denote the trained score network as
\begin{equation}
	{{\bf{s}}_{{{\bf{\theta }}^*}}}({\bf{x}}_i,{\sigma _i})={\nabla _{{\bf{x}}}}\log {p_{{\sigma _i}}}({\bf{x}}_i),
	\label{Eq:ncsn}
\end{equation}
where ${{\bf{\theta }}}$ represents the learnable parameters of the score network and the superscript ${^*}$ represents the trained parameters.
Reference \cite{meng2024generative} preliminarily explored the use of SGMs to generate modeling of seismic data. The results showed that SGMs can unconditionally generate rich and diverse seismic data, and the seismic data prior implicit in the trained score network can be directly applied to the posterior sampling of the seismic inverse problem\cite{meng2025posterior,meng2024stochastic}.

\subsubsection{Unconditional sampling of seismic data} 
In sampling stage (backward direction), \cite{song2019generative} run $M$ steps of Langevin Markov Chain Monte Carlo (MCMC) to get a sample for each  
${p_{{\sigma _i}}}({\bf{x}})$ sequentially: 
\begin{equation}
	{\bf{x}}_i^m \leftarrow {\bf{x}}_i^{m - 1} + {\alpha _i}{{\bf{s}}_{{{\bf{\theta }}^*}}}({\bf{x}}_i^{m - 1},{\sigma _i}) + \sqrt {2{\alpha _i}} {\bf{z}}_i^m, m = 1,2, \cdots M.
	\label{Eq:LD}
\end{equation}
where ${\alpha _i} = \varepsilon{{\sigma _i^2} \mathord{\left/{\vphantom {{\sigma _i^2} {\sigma _L^2}}} \right.\kern-\nulldelimiterspace} {\sigma _L^2}}$  is the step size,
$\varepsilon$  is learning rate and ${\bf{z}}_i^m$  is standard normal. The above is repeated for $i = 1,2, \cdots L$ , then the prior distribution transitions to the target distribution through all   noise scales (see reverse direction in Figure 1), 
${\bf{x}}_1^0 \sim {\mathcal N}({\bf{x}}|{\bf{0}},{\sigma_1}{\bf{I}})$,  
${\bf{x}}_L^M$  is the an exact sample from  ${p_{{\sigma _{\min }}}}({\bf{x}}) \approx {p_{{\rm{data }}}}({\bf{x}})$  ,so  ${\bf{x}} = {{\bf{x}}_L} + \sigma _L^2{{\bf{s}}_{\bf{\theta }}}({{\bf{x}}_L},{\sigma _L})$.

\subsection{Posterior sampling framework} 

Given an observation $\mathbf{y}$, its posterior distribution can be expressed as $p(\mathbf{x}|\mathbf{y})$, where  
\begin{equation} 
	\mathbf{y} = \mathbf{G}\mathbf{x} + \mathbf{n}, \label{eq:y} 
\end{equation}
$\mathbf{G}$ is the forward operator, and $\mathbf{n}$ represents measurement noise.
When the score function is replaced by a form that depends conditionally on $\mathbf{y}$, we can use Langevin dynamics to obtain a posterior solution that obeys the $p(\mathbf{x}|\mathbf{y})$
(see Figure 2)
\begin{equation}
	{\bf{x}}_i^m \leftarrow {\bf{x}}_i^{m - 1} + {\alpha _i}{\nabla _{{\bf{x}}_i^{m - 1}}}\log {p_{{\sigma _i}}}({\bf{x}}_i^{m - 1}|{\bf{y}}) + \sqrt {2{\alpha _i}} {\bf{z}}_i^m,m = 1,2, \cdots M,i = 1,2, \cdots ,L.
	\label{Eq:cLD}
\end{equation}
where ${\mathbf{x}}_i = {\mathbf{x}} + \mathbf{z}$ and $\mathbf{z} \sim \mathcal{N}\left(\mathbf{0}, \sigma_i^{2}\mathbf{I}\right)$, ${\nabla _{{\mathbf{x}}_i}}\log p_{\sigma _i}({\mathbf{x}}_i|{\bf{y}})$ is a conditional score function. Noted that, according to Bayes' theorem, the score function of $p(\mathbf{x}|\mathbf{y})$ can be written as: 
\begin{equation}
	{\nabla _{\bf{x}}}\log p({\bf{x}}|{\bf{y}}) = {\nabla _{\bf{x}}}\log p({\bf{x}}) + {\nabla _{\bf{x}}}\log p({\bf{y}}|{\bf{x}}),
	\label{Eq:bayes}
\end{equation}
where ${\nabla _{\bf{x}}}\log p({\bf{x}})$ can be approximated based on the trained score network ${{\mathbf{s}}_{{{\bf{\theta }}^*}}}({\mathbf{x}}_i,{\sigma _i})$, and ${\nabla _{\bf{x}}}\log p({\bf{y}}|{\bf{x}})$ is an an approximate analytical expression related to the forward operator and the noise distribution parameters. This means that the entire posterior sampling process via Equation \ref{Eq:LD}  does not need to retrain the score network and is unsupervised.

\subsubsection{Conditional score function}
Considering the scalability of forward operator $\mathbf{G}$ and unknown measurement noise $\mathbf{n}$, we use the conditional score function in the spectral space, and perform sampling in the SVD domain.  
Denote $\mathbf{G}=\mathbf{U} \mathbf{\Sigma} \mathbf{V}^T$, where $\mathbf{U} \in \mathbb{R}^{M \times M}$ and $\mathbf{V} \in \mathbb{R}^{N \times N}$ are orthogonal matrices, and $\mathbf{\Sigma} \in \mathbb{R}^{M \times N}$ is a rectangular diagonal matrix containing the singular values of $\mathbf{H}$, denoted as $\left\{s_j\right\}_{j=1}^{M}$ in descending order (${s_1 > s_2 > \dots > s_{M-1} > s_M \ge 0}$), $s_{j} = 0$ for $j = M+1, \dots, N$.
Notice that  $p\left({\mathbf{x}}_i | \mathbf{y}\right) = p\left({\mathbf{x}}_i | \mathbf{U}^T \mathbf{y}\right) = p\left(\mathbf{V}^T {\mathbf{x}}_i | \mathbf{U}^T \mathbf{y}\right)
$
since the multiplication of $\mathbf{y}$ by the orthogonal matrix $\mathbf{U}^T$ or the multiplication of $\mathbf{x}_i$ by $\mathbf{V}^T$ does not change its probability.
Thus score function  $\nabla_ {\mathbf{x}_i} \log p\left( {\mathbf{x}}_i |  \mathbf{y}\right)$ in pixel space is  equivalent with the score function  $\nabla_{\mathbf{V}^{T} {\mathbf{x}}_i} \log p\left(\mathbf{V}^{T} {\mathbf{x}}_i | \mathbf{U}^T \mathbf{y}\right)$ inspired by \cite{kawar2021snips}:
\begin{equation}
	\label{eqn:gradient}
	\nabla_{\mathbf{V}^T\mathbf{x}_i}\log p(\mathbf{V}^T\mathbf{x}_i|\mathbf{U}^T\mathbf{y})=
	\mathbf{\Sigma}^T|\sigma_\mathbf{y}^2\mathbf{I}-\sigma_i^2\mathbf{\Sigma\Sigma}^T|^{\dagger}
	(\mathbf{U}^T\mathbf{y}-\mathbf{\Sigma}\mathbf{V}^T\mathbf{x}_i)
	+\left.(\mathbf{V}^T\nabla_{\mathbf{V}^T\mathbf{x}_i}\log p(\mathbf{V}^T\mathbf{x}_i))\right|_{\leq},
\end{equation}
where the second term $\left. \left( \cdot \right) \right|{\leq}$ represents the vector $\cdot$ where the values at entries $j$ are retained if $s_j = 0$ or $0 < \sigma_i s_j < \sigma_\mathbf{y}$, and set to zero if $\sigma_i s_j > \sigma_\mathbf{y}$. The score function in the second term,
$\nabla_{\mathbf{V}^T\mathbf{x}_i}\log p(\mathbf{V}^T\mathbf{x}_i)$, is known and can be estimated using a pretrained score network $\mathbf{s}_{{\theta}^*}\left({\mathbf{x}}_i, \sigma_{i}\right)$ , which also introduces the  seismic-aware generative prior. Here, $\sigma_\mathbf{y}$ represents the noise level of the observation, which is a scalar to be estimated. 

\subsubsection{Posterior sampling using Langevin dynamics and conditional score function}
\begin{figure}[!htb]
	\centering
	\includegraphics[width=\textwidth]{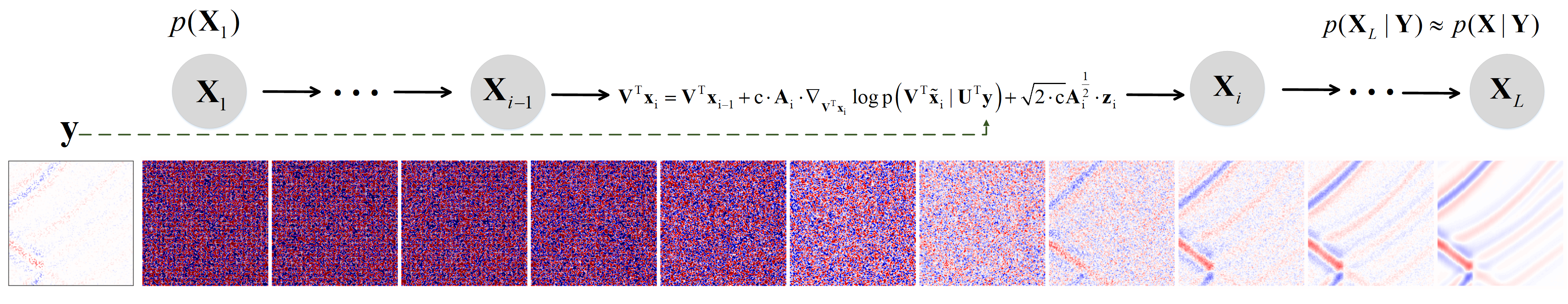}
	\caption{Schematic diagram of posterior sampling using Langevin dynamics and conditional score function. }
	\label{fig:x_y_evol}
\end{figure}

Thus, we can first sample a $\mathbf{V}^{T} {\mathbf{x}}_i$ from $p\left(\mathbf{V}^{T} {\mathbf{x}}_i | \mathbf{U}^T \mathbf{y}\right)$ 
using Langevin dynamics with conditional score function in spectral space, which can be described as
\begin{equation}
	\label{eqn:update_formula}
	\mathbf{V}^T \mathbf{x}_i = \mathbf{V}^T \mathbf{ x}_{i-1} + c \cdot \mathbf{A}_i \cdot \nabla_{\mathbf{V}^T \mathbf{x}_i} \log p\left(\mathbf{V}^T \mathbf{\tilde x}_i | \mathbf{U}^T\mathbf{y}\right) + \sqrt{2\cdot c}  \mathbf{A}_i^\frac{1}{2} \cdot \mathbf{z}_i,
\end{equation}
where $c$ is some constant,   ${\mathbf{A}_i = {diag}\left(\text{\boldmath$\alpha$}_i\right)}$ is the step size where $\text{\boldmath$\alpha$}_i \in \mathbb{R}^N$ can be represented as
\begin{equation}
	\label{eqn:update_step_size}
	\left(\text{\boldmath$\alpha$}_i\right)_j = \begin{cases}
		\sigma_i^2, & s_j = 0 \\
		\sigma_i^2 - \frac{\sigma_\mathbf{y}^2}{s_j^2}, & \sigma_{i} s_j > \sigma_\mathbf{y} \\
		\sigma_i^2 \cdot \left( 1 - s_j^2 \frac{\sigma_i^2}{\sigma_\mathbf{y}^2} \right), & 0 < \sigma_{i} s_j < \sigma_\mathbf{y},
	\end{cases}
\end{equation}  
then  multiplying $\mathbf{V}_T{\mathbf{x}}_i$ by $\mathbf{V}$ to obtain sample ${\mathbf{x}}_i$ in the pixel space, which is equivalent to sampling ${\mathbf{x}}_i$ from  $p\left({\mathbf{x}}_i | \mathbf{y}\right)$.  We present an illustration of the posterior sampling process in Figure \ref{fig:x_y_evol}, using the compressed sensing task as an example. 

In Sections \ref{Forward_operators} and \ref{Memory_Efficient_SVD}, the forward operator $\mathbf{G}$ in Equation \ref{eqn:update_formula} is introduced, and its low-memory SVD decomposition is presented. Furthermore, in Section \ref{Noise_level_estimation}, an automatic noise-level estimation algorithm is provided to estimate the parameter $\sigma_\mathbf{y}$ in Equation \ref{eqn:update_step_size}.

%  ${L_n}$ is an index of noise scale sequence such that ${L_n} \leftarrow \sigma ({\bf{y}}) \approx {\sigma _{{L_n}}} \in \left\{ {{\sigma _i}} \right\}_{i = 1}^L$,  $\sigma (\bf{y})$ represents the noise variance of the observation, ${\bf{s}}({\bf{y}}|{{\bf{x}}_i},{\sigma _i}) = ({\bf{y}} - {{\bf{x}}_i})/(\sigma _{{L_n}}^2 - \sigma _i^2)$ . 

We present the complete posterior sampling framework in Algorithm~\ref{alg:general}.
\begin{algorithm}[H]
	\caption{Posterior Sampling Framework}
	\label{alg:general}
	\begin{algorithmic}[1]
		\Require $\mathbf{y}$, $\mathbf{G}$, $\mathbf{s}_{{\theta}^*}, \left\{\sigma_{i}\right\}_{i=1}^{L}$, $c$, $\tau$ \Comment{$\mathbf{s}_{{\theta}^*}, \left\{\sigma_{i}\right\}_{i=1}^{L}$, $c$, $\tau$ is known and $\mathbf{s}_{{\theta}^*}$ is the pretrained score function of the distribution of seismic data }
		\State $\mathbf{U}, \mathbf{\Sigma}, \mathbf{V} \Leftarrow memory{\_}efficient{\_} svd\left(\mathbf{G}\right)$ \Comment{singular value decomposition of forward operator}
		%		\State $\mathbf{s}_{{\theta}^*}\left(\mathbf{x}_{t-1}, \sigma_{i}\right)$ \Comment{pretrained score function of prior distribution of seismic data}
		\State ${\sigma_\mathbf{y}} \Leftarrow median(\boldsymbol{\sigma}_\mathbf{y})$ \Comment{noise level estimation of seismic noise via Equation \ref{Eq:sigma_map}}
		\State Initialize $\mathbf{x_0}$ with random noise from $\mathcal{N}\left(\mathbf{0}, \mathbf{I}\right)$
		\If{$\mathbf{G} = \mathbf{I}$}
		\State $l \leftarrow \arg\min_i \left| \sigma_{\mathbf{y}} - \sigma_i \right|,
		\quad \text{where } {\sigma_l} \in \left\{ {\sigma_i} \right\}_{i=1}^L$ \Comment{$l$ is the starting index of sampling using Langevin dynamics}
		\Else
		\State $l \Leftarrow 1$
		\EndIf
		\For{$i \Leftarrow l$ to $L$}
		\State Obtain $\mathbf{A}_{i}$ via Equation \ref{eqn:update_step_size}
		\For{$t \Leftarrow 1$ to $\tau$}
		\State Draw $\mathbf{z}_{t} \sim \mathcal{N}\left(0, \mathbf{I}\right)$
		\State $\mathbf{d}_t \Leftarrow 
		\mathbf{\Sigma}^T \cdot \left| \sigma_\mathbf{y}^2 \mathbf{I} - \sigma_i^2 \mathbf{\Sigma \Sigma}^T \right|^{\dagger} \cdot
		\left( \mathbf{U}^T \mathbf{y} - \mathbf{\Sigma V}^T \mathbf{x}_{t-1} \right) +
		\left. \left( \mathbf{V}^T \cdot {{\bf{s}}_{{{\bf{\theta }}^*}}}\left(\mathbf{x}_{t-1}, \sigma_{i}\right) \right) \right|_{\leq}$
		\Comment{${{\bf{s}}_{{{\bf{\theta }}^*}}}$ is known}
		\State $\mathbf{x}_t \Leftarrow \mathbf{V} \cdot \left( \mathbf{V}^T \mathbf{x}_{t-1} + c \mathbf{A}_{i} \mathbf{d}_{t} + \sqrt{2c} \mathbf{A}_i^\frac{1}{2} \mathbf{z}_t \right)$
		\EndFor
		\State $\mathbf{x}_0 \Leftarrow \mathbf{x}_\tau$
		\EndFor
		\Ensure $\mathbf{x_0}$
	\end{algorithmic}
\end{algorithm}

\subsubsection{Forward operators of different tasks}\label{Forward_operators}
\begin{figure}[!htb]
	\centering
	\includegraphics[width=\textwidth]{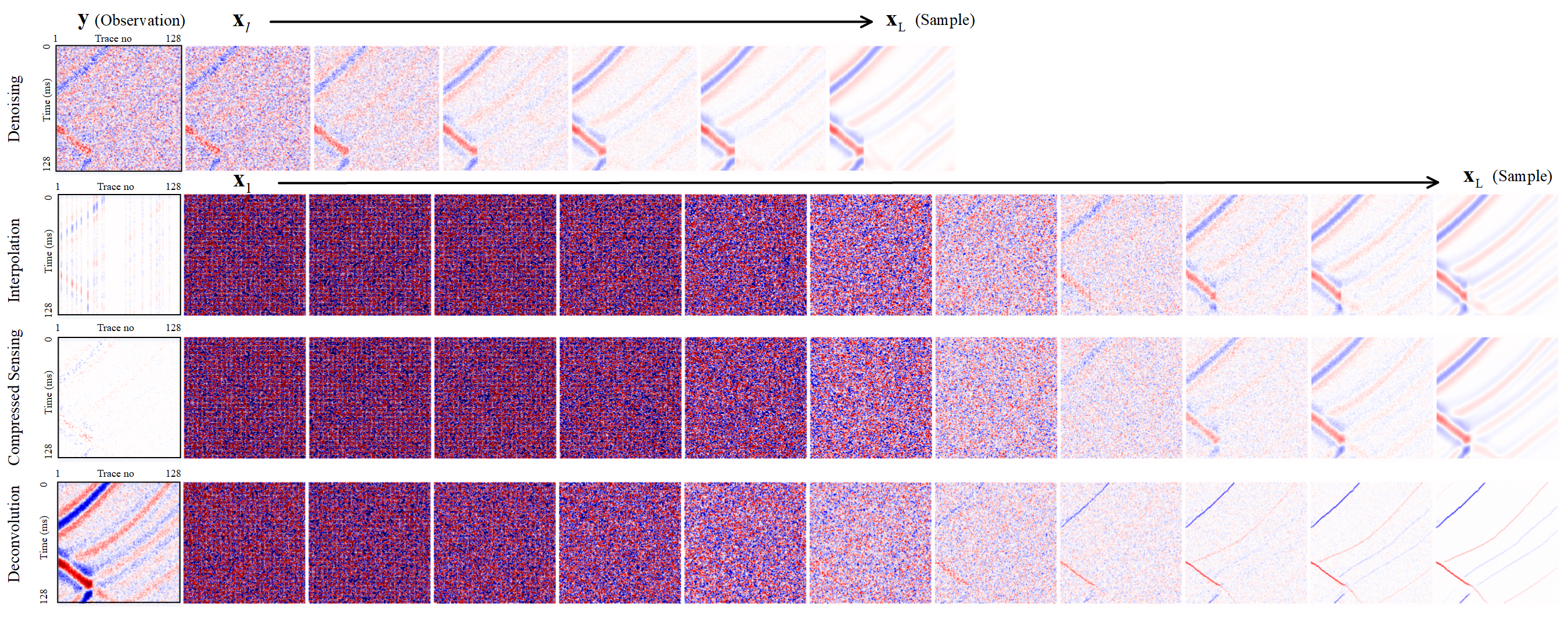}
	\caption{Schematic diagram of the posterior sampling trajectory of observations for different degradation processes}
	\label{fig:x_y_evol_more}
\end{figure}

For the denoising task, we set $\mathbf{G} = \mathbf{I}$, where $\mathbf{I}$ is the identity matrix.  
For the interpolation task, $\mathbf{G} = \mathbf{M}$, where $\mathbf{M}$ is a binary mask matrix indicating the positions of missing entries.  
For the compressed sensing task, $\mathbf{G} = \mathbf{P} \mathbf{W}$, where $\mathbf{P} \in \mathbb{R}^{d_0 \times d}$ is a random sampling operator and $\mathbf{W} \in \mathbb{R}^{d \times d}$ is a transformation matrix. The observation model is expressed as
$ 
\mathbf{y} =\mathbf{G} \mathbf{x} + \mathbf{n}= \mathbf{P} \mathbf{W} \mathbf{x} + \mathbf{n}, $
where $\mathbf{y} \in \mathbb{R}^{d_0}$ and $d_0 < d$, with $\mathbf{n}$ denoting additive noise, ${d_0}/d$ is is the compression ratio.
For deconvolution task, $\mathbf{G}=\mathbf{W}$, where $\mathbf{W}$ is a Toeplitz matrix formed by the seismic wavelet. 
Figure \ref{fig:x_y_evol_more} illustrates the posterior sampling schematic for inverse problems corresponding to different forward operators.
Note that when $\mathbf{G} = \mathbf{0}$ and $\sigma_{\mathbf{y}}=0$,  implying no measurements, the posterior sampling algorithm will degenerate into unconditional sampling, that is, signal generation. 

Note that, if the task is denoising, 
the posterior sampling trajectory starts from the $p({\bf{x}}_{l})$ distribution (see Figure \ref{fig:x_y_evol_more}) so that ${{\bf{x}}_{l}} \approx {\bf{y}}$, $l$ is an index of noise scale sequence such that  
\begin{equation}
	l \leftarrow \arg\min_i \left| \sigma_{\mathbf{y}} - \sigma_i \right|,
	\quad \text{where } {\sigma_l} \in \left\{ {\sigma_i} \right\}_{i=1}^L,
	\label{Eq:sigma_est}
\end{equation}

\begin{figure}[htb!]
	\setlength{\abovecaptionskip}{0.3cm}
	\centering
	% 第一行 (a)
	\subfloat[]{\includegraphics[width=\textwidth]{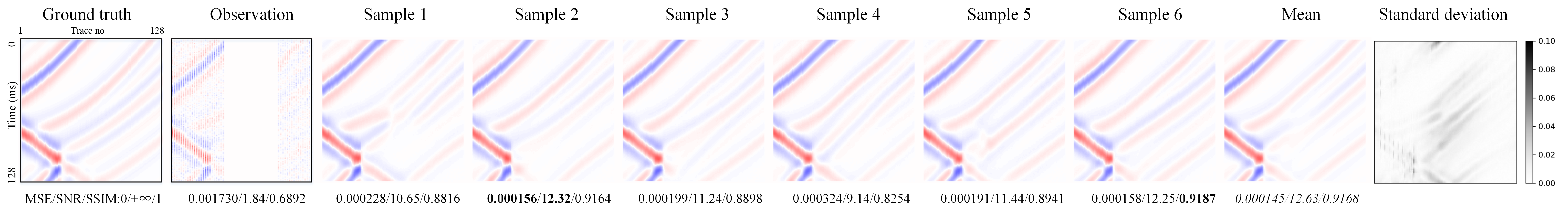}
		\label{fig:sample_compare_a}}\\[0.4cm]
	% 第二行 (b)
	\subfloat[]{\includegraphics[width=\textwidth]{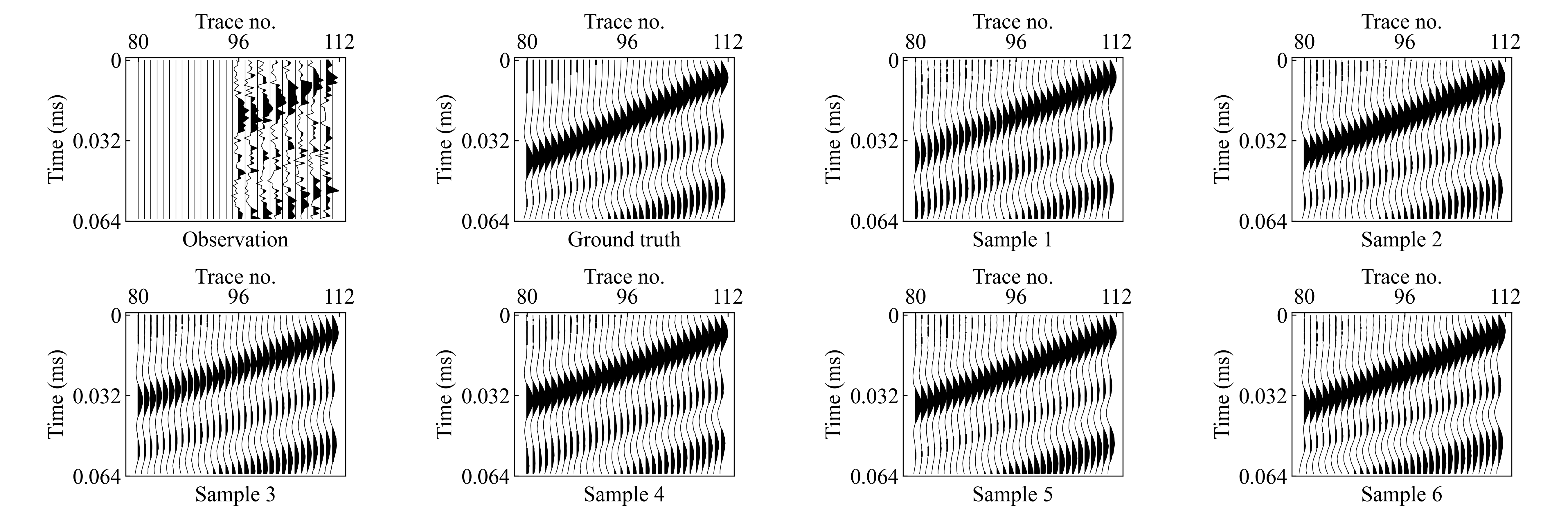}
		\label{fig:sample_compare_b}}
	\caption{Schematic diagram of multiple posterior solutions and their mean and standard deviation, (a) original image (b) locally enlarged wiggle image}
	\label{fig:sample_compare}
\end{figure}

Figure \ref{fig:sample_compare_a}, taking the ill-posed interpolation task as an example, presents multiple stochastic solutions sampled from the posterior distribution. The “mean” and “standard deviation” represent the mean (i.e., the expectation of the posterior distribution) and the standard deviation of the sampled solutions. From both visual effects and numerical evaluation, it can be observed that the sampled solutions are diverse and of high quality, where solutions with higher numerical values are closer to the ground truth, as shown in the detailed local zoom-in view in Figure \ref{fig:sample_compare_b}. The mean and standard deviation further facilitate uncertainty analysis.

%\begin{figure}[!htb]
%	\centering
%	\includegraphics[width=\textwidth]{images/sample_compare.png}
%	\caption{Schematic diagram of a score-based generative model. }
%	\label{fig:sample_compare}
%\end{figure}

\subsection{Memory Efficient SVD}\label{Memory_Efficient_SVD}
Due to the storage of the the matrix $\mathbf{V}$, the above posterior sampling algorithm has a space complexity of $\mathcal{O}(d^2)$ for signals of size $d$, making it impossible to process large-scale seismic data. Inspired by \cite{kawar2022denoising}, we use a memory-efficient SVD decomposition to reduce the complexity to $\mathcal{O}(d)$ by leveraging special properties of the matrices $\mathbf{G}$ used.  Table \ref{tab:complexity} compares the proposed PSF method with several representative SGM-based Posterior Sampling methods, including SGMPS-RNA \cite{meng2025posterior}, SGMPS-SDR \cite{meng2024stochastic}, and SNIPS \cite{kawar2021snips}, where SGMPS-RNA is designed for random noise attenuation, and SGMPS-SDR focuses on simultaneous denoising and reconstruction. However, these task-specific methods are typically restricted to their respective degradation types.
In contrast, the proposed PSF performs multi-purpose posterior sampling in the spectral domain, effectively handling various noisy inverse problems within a unified framework.
Notably, when the forward operator satisfies $\mathbf{G} = \mathbf{I}$, PSF degenerates to SGM-RNA; when $\mathbf{G} = \mathbf{M}$, it degenerates to SGM-SDR; and when $\mathbf{G} = \mathbf{0}$ and $\mathbf{n} \sim p(\mathbf{x}_1)$, PSF naturally reduces to unconditional sampling.
Moreover, PSF maintains the same linear space complexity $\mathcal{O}(d)$ as other sampling methods, while avoiding the quadratic cost $\mathcal{O}(d^2)$ of SNIPS\cite{kawar2021snips}.
This demonstrates that PSF achieves a favorable balance between task generality and computational efficiency, offering both flexibility and scalability for practical seismic restoration tasks.

Table \ref{tab:speed} compares the inference speed and memory performance of different posterior sampling algorithms under increasing data dimensions. NFEs/s denotes the number of function evaluations (NFEs) per second, which measures the sampling efficiency by indicating how many times the neural network (or score function) is evaluated within one second during the sampling process.
SNIPS exhibits severe memory constraints when processing higher-dimensional seismic data, resulting in out-of-memory errors regardless of whether the score function is implemented by a lightweight UNet\cite{ronneberger2015u} or a heavier Noise Conditional Score Network(NCSNv2)\cite{song2020improved} architecture.
In contrast, PSF maintains stable inference over higher-dimensional data, benefiting from its Memory-Efficient SVD formulation, which substantially reduces spatial complexity and enables scalability to large data dimensions.
The parentheses in the Method column indicate the specific score-function backbone used in each case.
By default, SNIPS\cite{kawar2021snips} employs NCSNv2 as its unconditional score model; for fair comparison, all models were trained under identical conditions using the same seismic dataset.

%\begin{table}[htbp]
%	\centering
%	\caption{Space complexity of different posterior sampling method via Langevin dynamic ($d$ is data dimension).}
%	\label{tab:complexity}
%	\begin{tabular}{lcccc}
%		\toprule
%		Methods & PSF(proposed) & SGMPS-RNA\cite{meng2025posterior} & SGMPS-SDR\cite{meng2024stochastic} & SNIPS\cite{kawar2021snips}\\
%		\midrule
%		Multi-purpose posterior sampling & $\checkmark$ & $\times$ & $\times$ 
%		& $\checkmark$  \\
%		%		\midrule
%		Sampling domain & spectral & pixel & pixel 
%		& spectral  \\
%		Space complexity & $\mathcal{O}(d)$ & $\mathcal{O}(d)$ & $\mathcal{O}(d)$ & $\mathcal{O}(d^2)$ \\
%		\bottomrule
%	\end{tabular}
%\end{table}

\begin{table}[htbp]
	\centering
	\caption{The space complexity of different posterior sampling method via Langevin dynamic ($d$ is data dimension).}
	\label{tab:complexity}
	\begin{tabular}{l c c c c}
		\toprule
		Methods & \makecell{PSF\\ (proposed)} & \makecell{SGMPS-RNA\\\cite{meng2025posterior}} & \makecell{SGMPS-SDR\\\cite{meng2024stochastic}} & \makecell{SNIPS\\\cite{kawar2021snips}}\\
		\midrule
		\makecell{Multi-purpose\\ posterior sampling} & $\checkmark$ & $\times$ & $\times$ & $\checkmark$  \\
		%		Noise level estimation& $\checkmark$ & $checkmark$ & $checkmark$ & $\times$  \\
		Space complexity & $\mathcal{O}(d)$ & $\mathcal{O}(d)$ & $\mathcal{O}(d)$ & $\mathcal{O}(d^2)$ \\
		Sampling domain & spectral & pixel & pixel & spectral  \\
		\bottomrule
	\end{tabular}
\end{table}

\begin{table}[htbp]
	\centering
	\caption{Inference-speed comparison, taking an interpolation task with 50\% random missing seismic traces as an example}
	\label{tab:speed}
	\begin{tabular}{@{}lllrrl@{}}
		\toprule
		Method & $d$ & Time $\downarrow$ & NFEs/s $\uparrow$ & Note \\
		\midrule
		PSF (UNet)        & $128 \times 128$ & 13.32s & 41.66it/s  & -- \\
		PSF (NCSNv2)      & $128 \times 128$ & 60.79s & 8.71it/s  & -- \\
		SNIPS (UNet)      & $128 \times 128$ & 46.86s &  11.27it/s  & -- \\
		\makecell{SNIPS (NCSNv2)\\\cite{kawar2021snips}}    & $128 \times 128$ & 86.07s & 6.05it/s  & -- \\
		\midrule
		PSF (UNet)        & $256 \times 256$ & 20.21s & 30.93it/s  & -- \\
		PSF (NCSNv2)      & $256 \times 256$ & 170.50s & 3.05it/s & -- \\
		SNIPS (UNet)      & $256 \times 256$ & -- & -- & \multicolumn{2}{c}{Out of memory} \\
		\makecell{SNIPS (NCSNv2)\\\cite{kawar2021snips}}    & $256 \times 256$ &-- & -- & \multicolumn{2}{c}{Out of memory} \\
		\bottomrule
	\end{tabular}
\end{table}

\subsection{Noise level estimation of seismic noise}\label{Noise_level_estimation}
The posterior sampling in Algorithm \ref{alg:general} requires an estimate of the noise level $\sigma_{\mathbf{y}}$,  which is a scalar.  
We employ a variational inference model, termed Variational Inference non-independent and non-identically distributed (VI-non-IID), trained on unlabeled synthetic seismic data \cite{meng2022seismic}, to directly predict the noise variance $\sigma_\mathbf{y}$. We directly utilize the open-source VI-non-IID model from \cite{meng2022seismic}.
Unlike conventional i.i.d. noise assumptions, VI-non-IID models the unknown noise distribution as a non-IID and pixel-wise Gaussian distribution~\cite{yue2019variational}:
\begin{equation}
	y_i \sim \mathcal{N}(z_i, \sigma_i^2), \quad i=1,2,\cdots,d,
	\label{Eq:generate_noisy_noniid}
\end{equation}
where $\mathbf{z}$ represents the latent clean data and $\sigma_i^2$ vary across pixels.
The variational approximate posterior of the noise variance, $q(\boldsymbol{\sigma}^2|\mathbf{y})$, is modeled as an inverse Gamma distribution:
\begin{equation}
	q(\boldsymbol{\sigma}^2|\boldsymbol{y}) = \prod_{i=1}^d \text{IG}(\sigma_i^2|\lambda_i, \zeta_i),
\end{equation}
with parameters $\boldsymbol{\lambda}, \boldsymbol{\zeta}$ predicted by a noise level estimation network (NLE-Net):
\begin{equation}
	\left[ \boldsymbol{\lambda}, \boldsymbol{\zeta} \right] = f_{\text{NLE}}(\boldsymbol{y};\theta_{\text{NLE}}).
	\label{Eq:sigma_output}
\end{equation}

Finally, the pixel-wise noise level is obtained as the mode of the inverse Gamma distribution:
\begin{equation}
	\boldsymbol{\sigma}_\mathbf{y}=\frac{\boldsymbol{\zeta}}{\boldsymbol{\lambda}+\boldsymbol{1}},
	\label{Eq:sigma_map}
\end{equation}
which yields a noise variance map $\boldsymbol{\sigma}_\mathbf{y}$ of the same shape as $\mathbf{y}$.
For stationary noise,
$\sigma_\mathbf{y}$ can be set as 
$median(\boldsymbol{\sigma}_\mathbf{y})$, for non-stationary noise, $\sigma_\mathbf{y}$ can be interactively selected within the interval $[min(\boldsymbol{\sigma}_\mathbf{y}),max(\boldsymbol{\sigma}_\mathbf{y})]$.  For an example of the pixel-wise noise level map and the effect of the input noise level on posterior sampling, please refer to Section~\ref{Examples}.

\section{Examples}\label{Examples}

%\subsection{Experimental setup}

\subsection{Pretrained score function and unconditional sampling}
\begin{figure}[!htb]
	%	\vspace{-2mm}
	\centering
	\includegraphics[width=\textwidth]{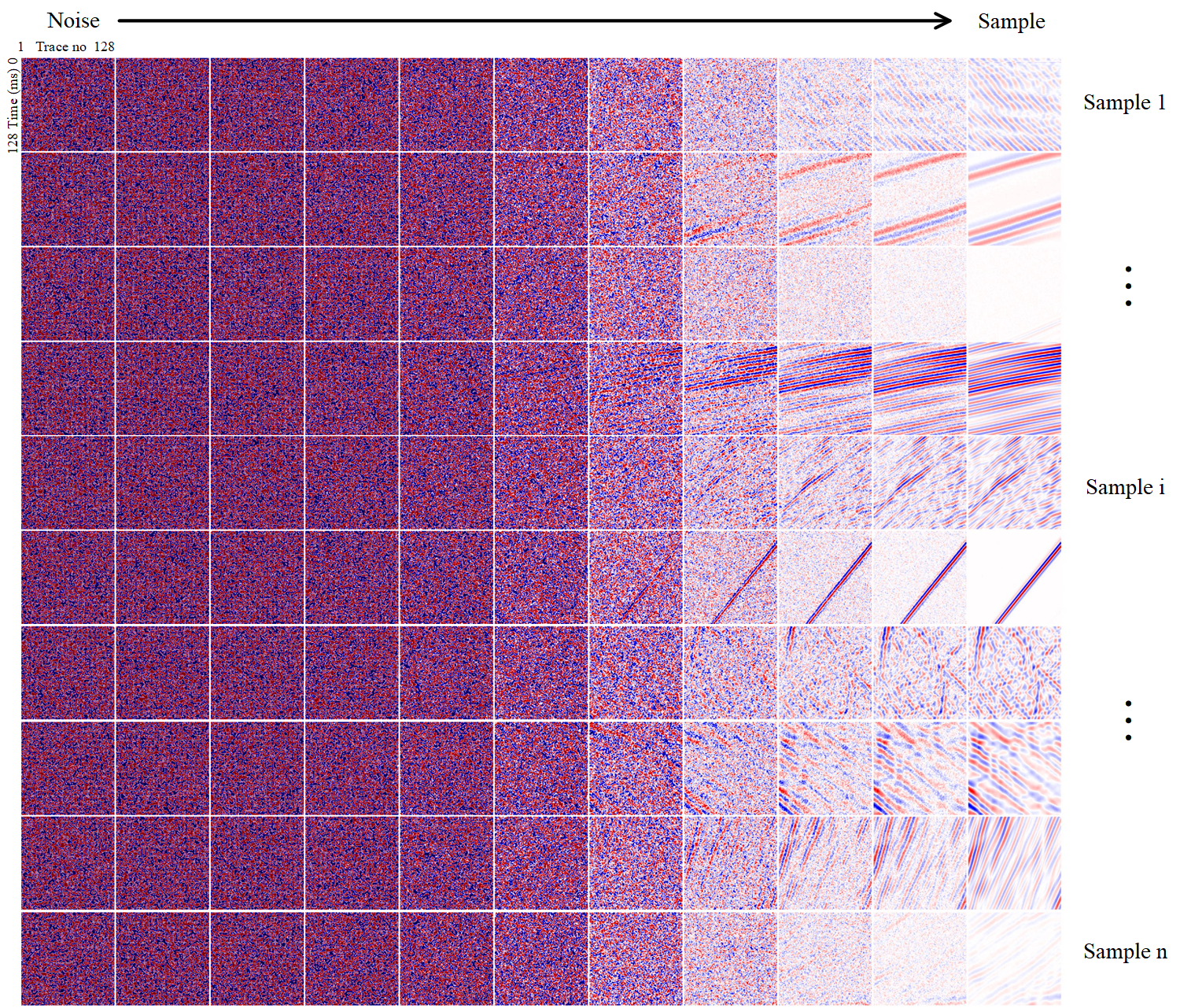}
	\caption{Schematic diagram of unconditional sampling using pretrained score function}
	\label{fig:unconditional_sampling}
	%	\vspace{-2mm}
\end{figure}

The pre-trained score function (${{\bf{s}}_{{{\bf{\theta }}^*}}}({\bf{x}}_i,{\sigma _i})$ in Equation \ref{Eq:ncsn}) is key to our method's posterior sampling. We directly used the open-source SGMs model (named SGM\_seismic) from Reference \cite{meng2024generative}, trained with unlabeled synthetic seismic data, for generative modeling (unconditional sampling) of seismic data. This not only validates the applicability of our method, which leverages an existing trained unconditional generative model, but also demonstrates that the entire process does not require ground truth labels, making it unsupervised. Figure \ref{fig:unconditional_sampling} shows seismic data sampled unconditionally using SGM. It can be observed that the samples have rich diversity in terms of linearity, curves, amplitudes, and inclinations. This shows that SGMs implies more complex seismic data priors that are difficult to express using general assumptions, rather than simple data prior assumptions such as linearity, low rank, and predictability.

The training hyperparameters for the two unconditional SGMs models used in the experiments are listed in Table \ref{tab:para}. SGM\_seismic is employed for general-purpose seismic data recovery tasks, while SGM\_reflection is specifically used for deconvolution. Since no publicly available unconditional generative model for reflection coefficients exists, we trained SGM\_reflection ourselves using synthetic data generated from the Marmousi model and the reflection coefficient dataset provided in \cite{yu2025unsupervised}.

\begin{table}[h]
	\centering
	\caption {Parameters involved in pre-training unconditional generative models in the  experiment}
	\label{tab:para}
	\begin{tabular}{lccccccc}
		\toprule
		Trained unconditional SGMs &	Data type & $c$ & $\tau$ & $L$ & $\sigma_1$ & $\sigma_L$ & $\frac{\sigma_{i+1}}{\sigma_i}$ \\
		\hline
		SGM\_seismic&	Seismic data & $3.3e-6$ & $5$ & $500$ & $32$ & $0.01$ & $0.983$ \\
		SGM\_reflection&	Reflection coefficient
		& $3.3e-6$ & $5$ & $500$ & $30$ & $0.01$ & $0.988$ \\
		\bottomrule
	\end{tabular}
\end{table}

Note that the hyperparameters $\left\{\sigma_{i}\right\}_{i=1}^{L}$, $c$, and $\tau$ in Algorithm \ref{alg:general} are fixed during sampling and require no tuning, since they are inherited from the training setup of the unconditional SGMs. The only task-related hyperparameter is the forward operator $\mathbf{G}$, which is determined based on the specific problem.

\subsection{Generalization out of distribution}

\begin{figure}[htb!]
	\setlength{\abovecaptionskip}{0.3cm}
	\centering
	% 第一行 (a)
	\subfloat[]{\includegraphics[width=0.5\textwidth]{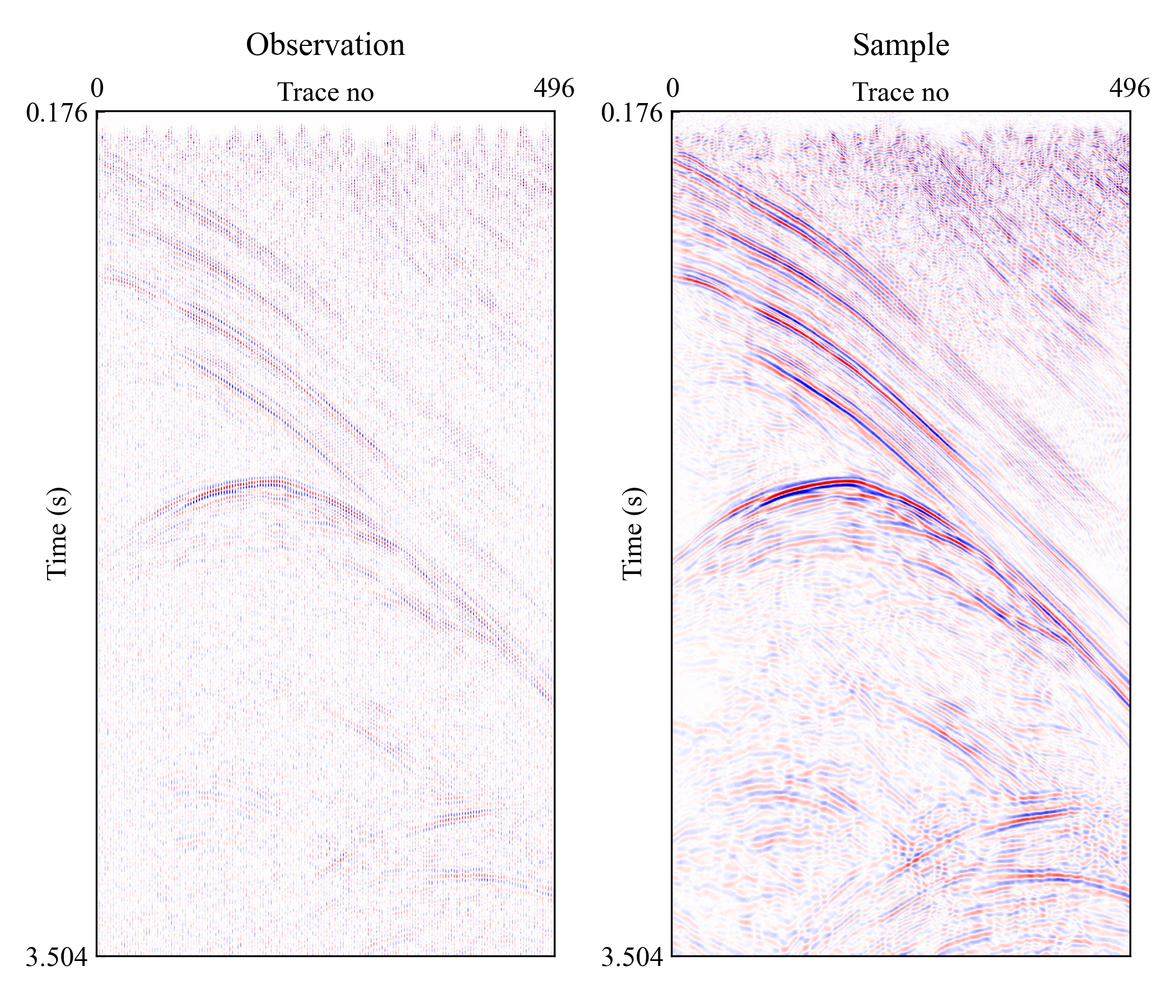}
		\label{fig:ood_a}}
	\subfloat[]{\includegraphics[width=0.5\textwidth]{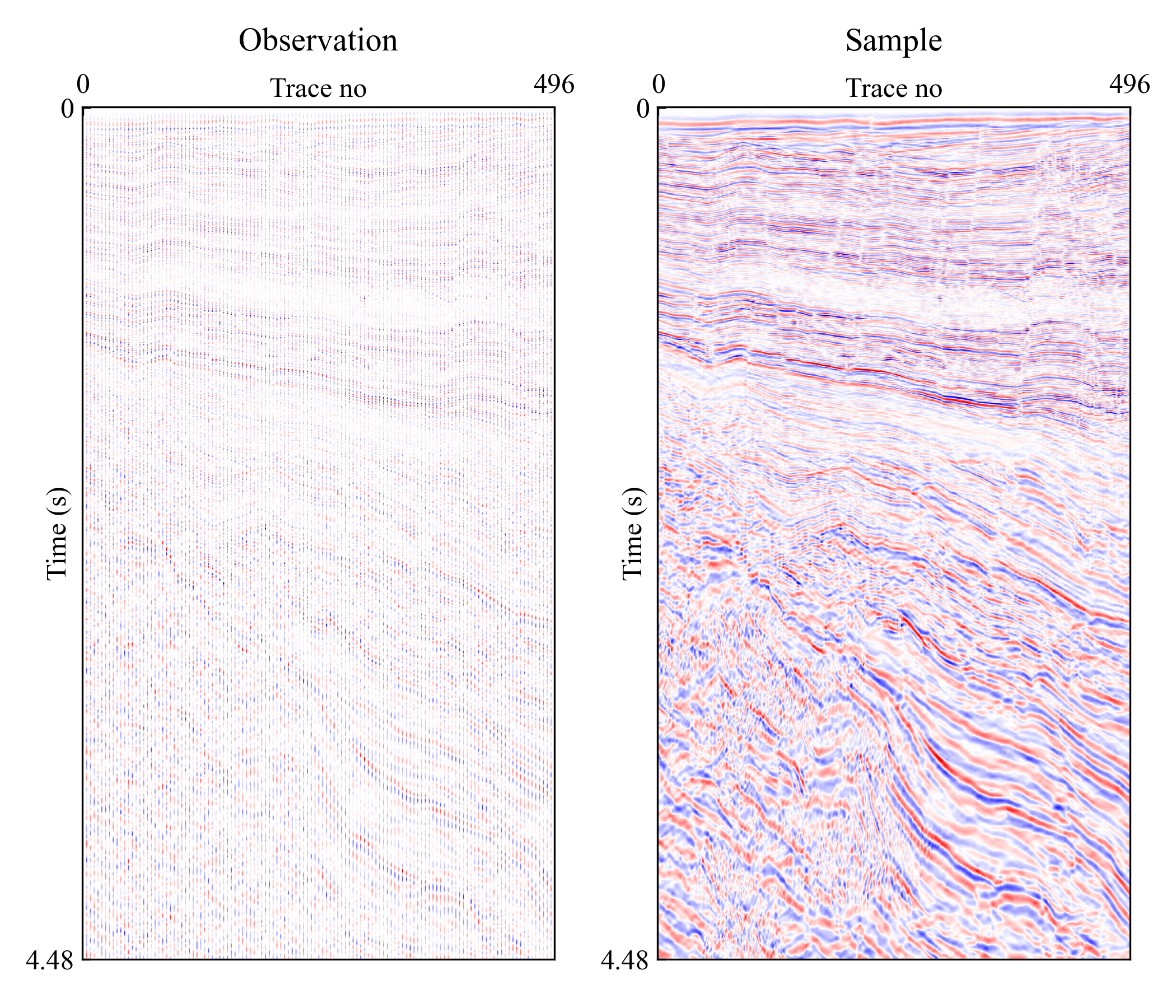}
		\label{fig:ood_b}}\\[-0.2cm]
	% 第二行 (b)
	\subfloat[]{\includegraphics[width=0.5\textwidth]{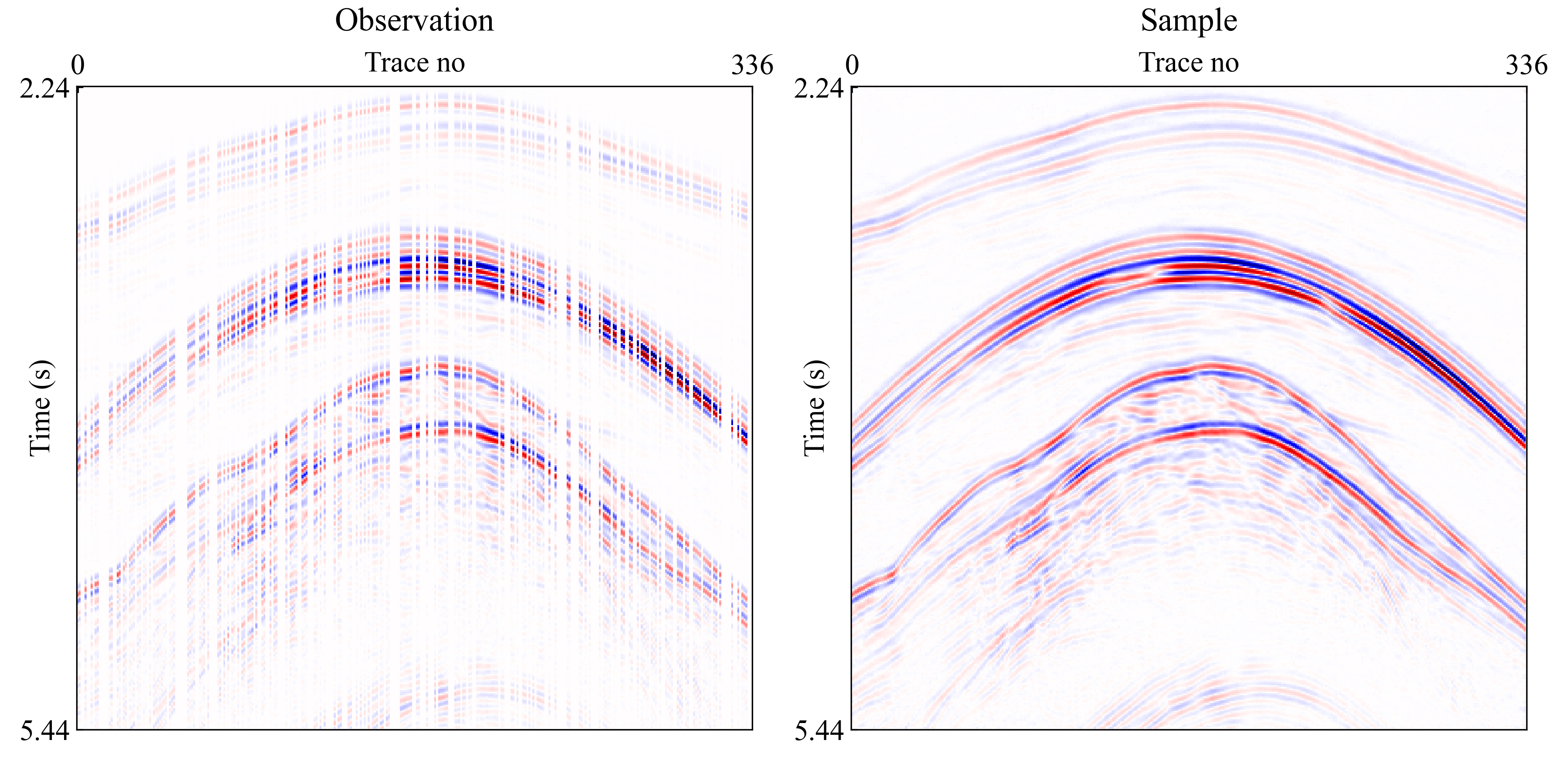}
		\label{fig:ood_c}}
	\subfloat[]{\includegraphics[width=0.5\textwidth]{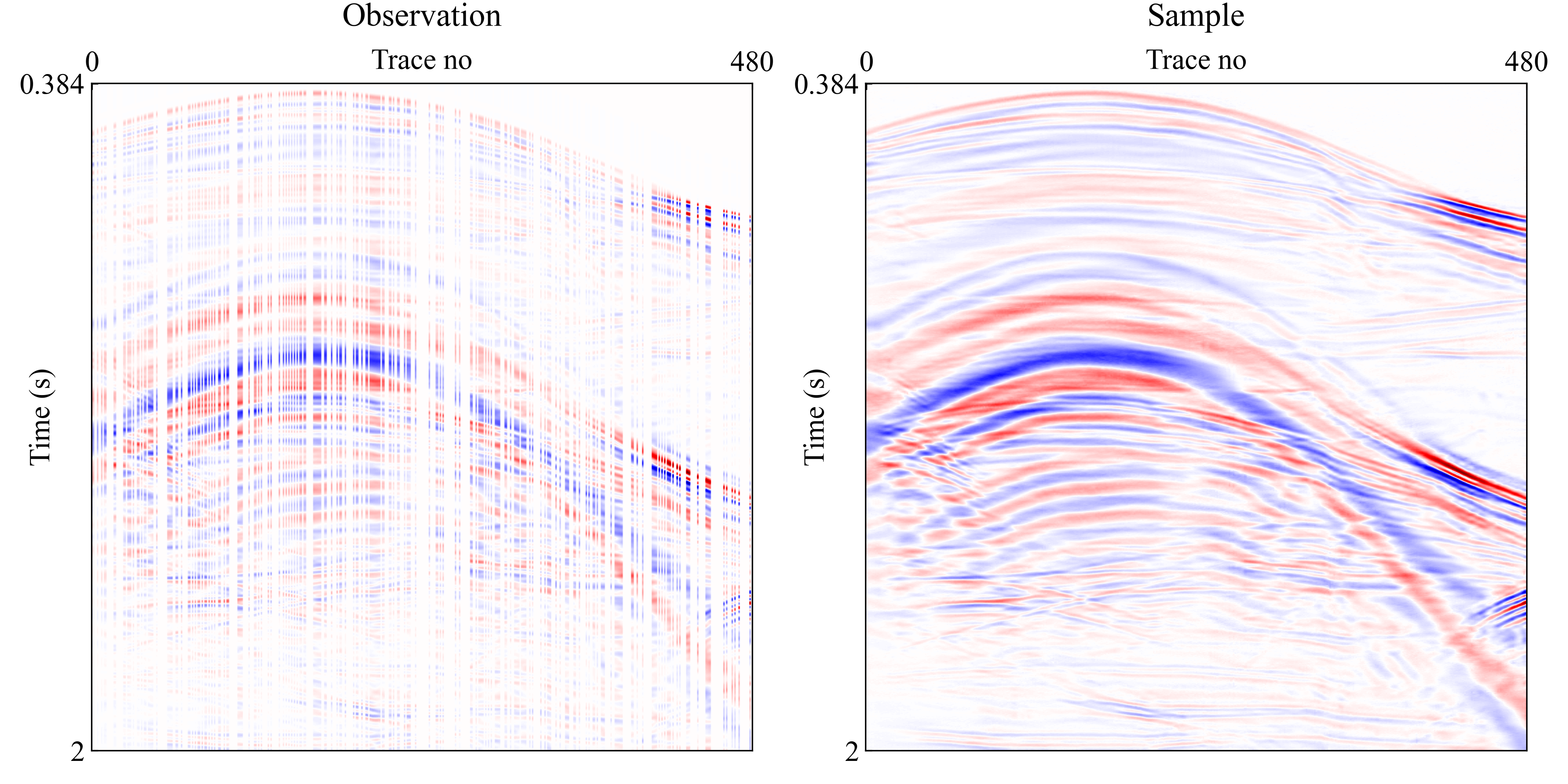}
		\label{fig:ood_d}}

	\caption{Out-of-distribution generalization results, taking seismic data interpolation as an example. (a,b) post-stack data , (c,d) prestack data.}
	\label{fig:ood}
\end{figure}

\begin{figure}[!htb]
	%	\vspace{-2mm}
	\centering
	\includegraphics[width=\textwidth]{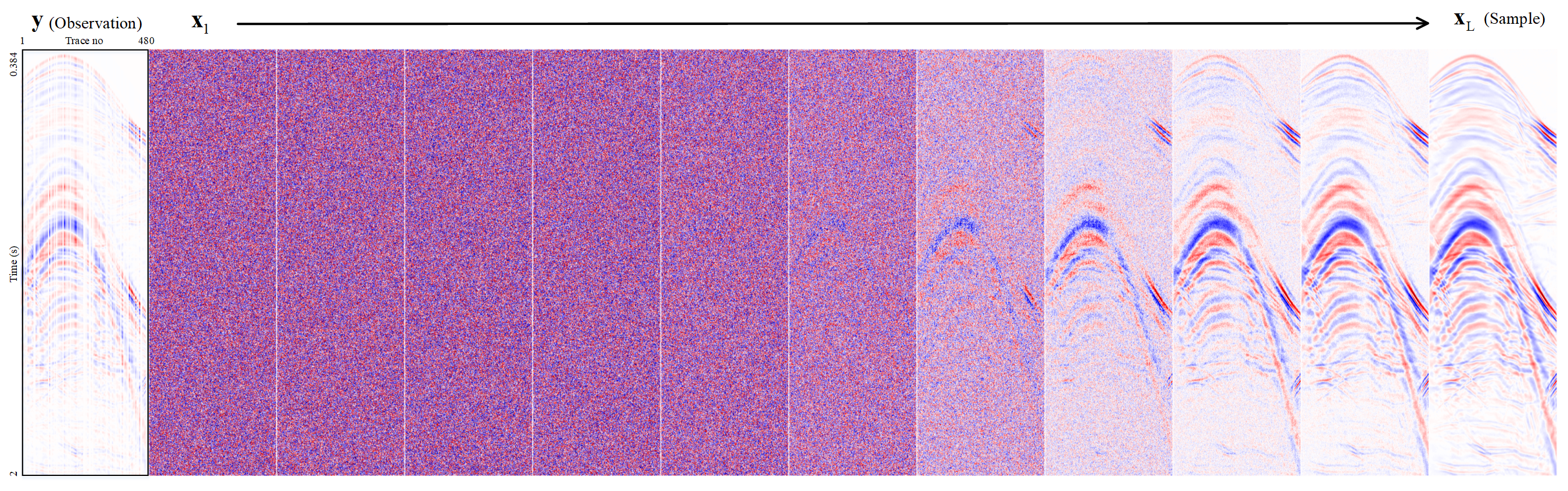}
	\caption{Schematic diagram of posterior sampling trajectory of the field data (in Figure \ref{fig:ood_d}) using PSF}
	\label{fig:XD_L3_evol}
	%	\vspace{-2mm}
\end{figure}

Because the score network pre-trained on synthetic data implicitly incorporates rich generative seismic data priors, it easily generalizes to unseen real data. We use a specific conditional generation task (seismic data interpolation) as an example to analyze the potential of generative seismic data priors for out-of-distribution generalization. Figure \ref{fig:ood} shows interpolation results for real data of different types (pre-stack, post-stack), from different regions, and with different missing patterns. The results show that the PSF consistently samples high-quality posterior solutions. Figure \ref{fig:XD_L3_evol} shows the posterior sampling trajectory of the field data in Figure \ref{fig:ood_d}, evolving from a Gaussian distribution to a posterior distribution. It can be observed that recovered data can be sampled from a posterior distribution conditional on the observed data. Figure 4 also shows that although the SGMs is generatively modeled on samples of a fixed size of 128*128, it can generate stochastic solutions of different sizes conditionally dependent on the observed data during the posterior sampling phase, which also illustrates the ability of the generative seismic data prior to generalize outside the distribution.

%\begin{figure}[htb!]
%	\setlength{\abovecaptionskip}{0.3cm}
%	\centering
%	\subfloat[]{\includegraphics[width=\textwidth]{images/XJ_den_v2.png}
%		\label{fig:xj_comp_a}}\\
%	\subfloat[]{\includegraphics[width=\textwidth]{images/XJ_int_v2.png}
%		\label{fig:xj_compb}}
%	\caption{Schematic diagram of multiple posterior solutions (Part 1)}
%	\label{fig:xj_comp_1}
%\end{figure}
%
%% 第二页显示后两个子图
%\begin{figure}[htb!]
%	\setlength{\abovecaptionskip}{0.3cm}
%	\centering
%	\subfloat[]{\includegraphics[width=\textwidth]{images/XJ_cs_v2.png}
%		\label{fig:xj_comp_c}}\\
%	\subfloat[]{\includegraphics[width=\textwidth]{images/XJ_dec_v2.png}
%		\label{fig:xj_comp_d}}
%	\caption{Schematic diagram of multiple posterior solutions (Part 2)}
%	\label{fig:xj_comp_2}
%\end{figure}

\subsection{Unsupervised posterior sampling}
PSF can unsupervisedly restore seismic data with unknown noise levels and different degradation types, making it flexibly applicable to different tasks.

\subsubsection{Denoising}
\begin{figure}[!htb]
	%	\vspace{-2mm}
	\centering
	\includegraphics[width=\textwidth]{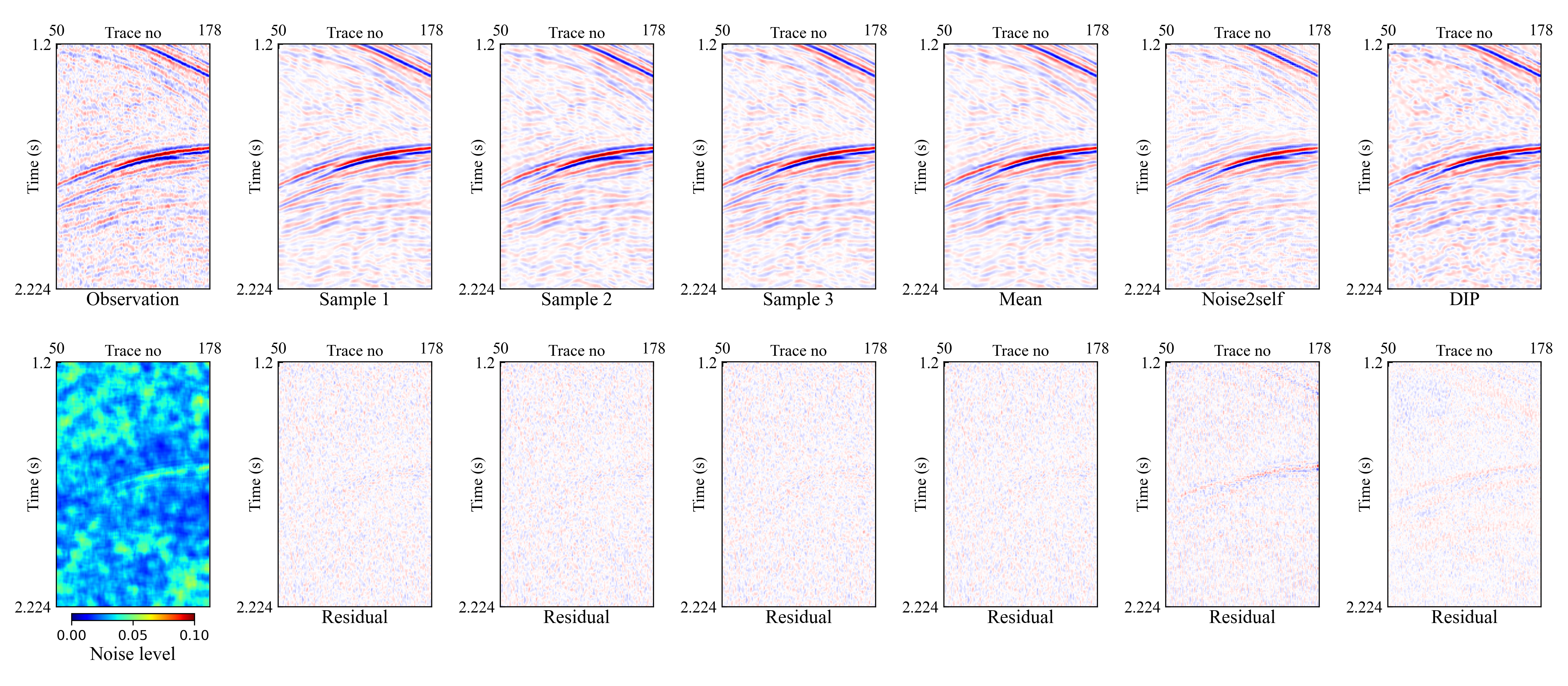}
	\caption{Unsupervised posterior sampling results, taking the denoising task as an example}
	\label{fig:XJ_den}
	%	\vspace{-2mm}
\end{figure}

Figure \ref{fig:XJ_den} shows the denoising results of different methods. Samples 1-3 are the results of three random samplings, and Mean represents the mean of the sample. The noise level map ($\boldsymbol{\sigma}_\mathbf{y}$) in Figure \ref{fig:XJ_den} is the pixel-wise noise level predicted using equation \ref{Eq:sigma_map}. We take $median(\boldsymbol{\sigma}_\mathbf{y})$ as the value of the input noise level ${\sigma_\mathbf{y}}$ in Algorithm \ref{alg:general}. 
We use  the self-supervised learning method Noise2Self\cite{batson2019noise2self} and the unsupervised deep learning method Deep Image Prior (DIP)\cite{ulyanov2018deep}  as baseline methods. 
For noise2self, we use a trace-wise mask training strategy instead of a pixel-wise mask training strategy, which is more consistent with the non-pixelwise independent characteristics of field noise and has been proven to be more effective for seismic data denoising\cite{meng2025self,fang2022bsnet}.The number of DIP iterations is set to 5000.
Compared with Noise2Self and DIP,
PSF can effectively suppresses random noise while minimizing the leakage of useful signals.

\subsubsection{Interpolation}
\begin{figure}[!htb]
	%	\vspace{-2mm}
	\centering
	\includegraphics[width=\textwidth]{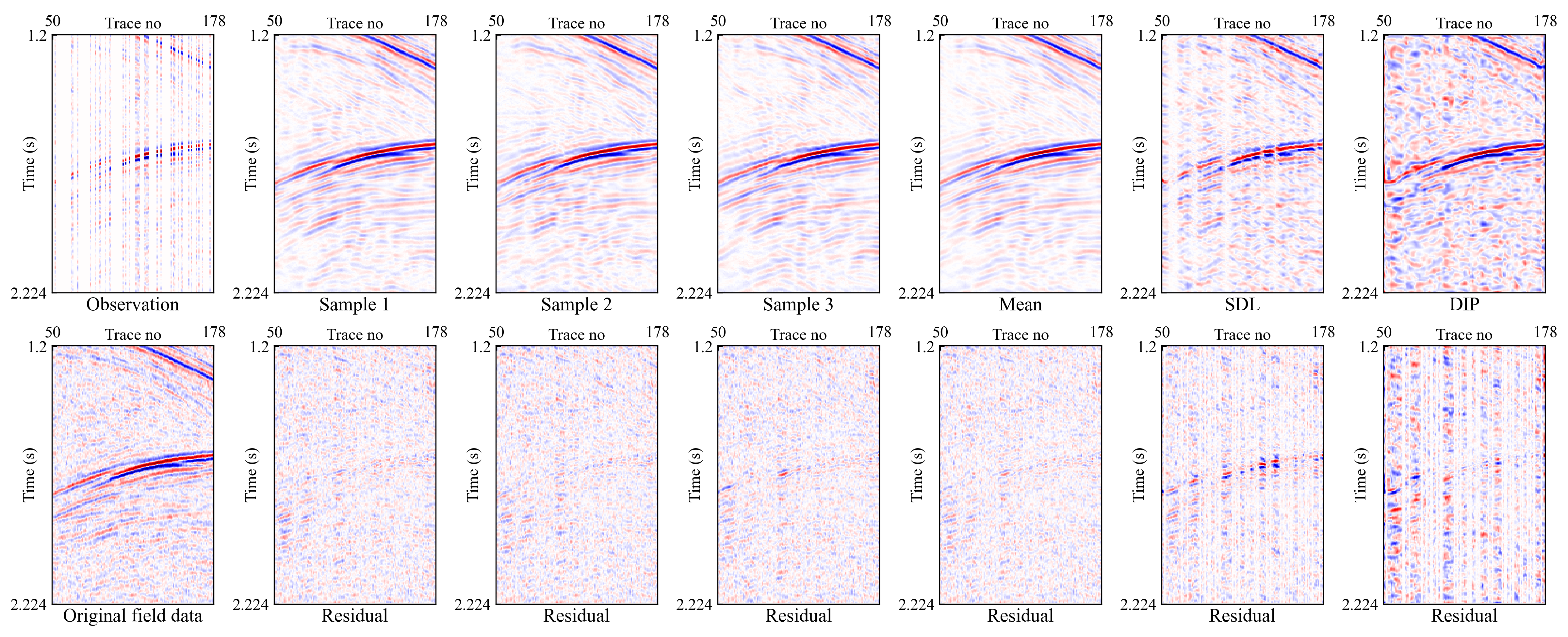}
	\caption{Unsupervised posterior sampling results, taking the interpolation task as an example}
	\label{fig:XJ_int}
	%	\vspace{-2mm}
\end{figure}

Figure \ref{fig:XJ_int} shows the seismic data interpolation results of different methods under the condition of 75\% random missing probability. We use supervised deep learning method(SDL) and DIP\cite{ulyanov2018deep} as the baseline method. The network used by SDL is UNet\cite{ronneberger2015u}, and the training data is the same as the data used to train SGM. For training details of SDL, please refer to \cite{yu2019deep}. PSF can simultaneously denoise and interpolate seismic data. The residual between the stochastic solution sampled and the original field data contains more noise and less useful signal.

\subsubsection{Compressed Sensing}
\begin{figure}[!htb]
	%	\vspace{-2mm}
	\centering
	\includegraphics[width=\textwidth]{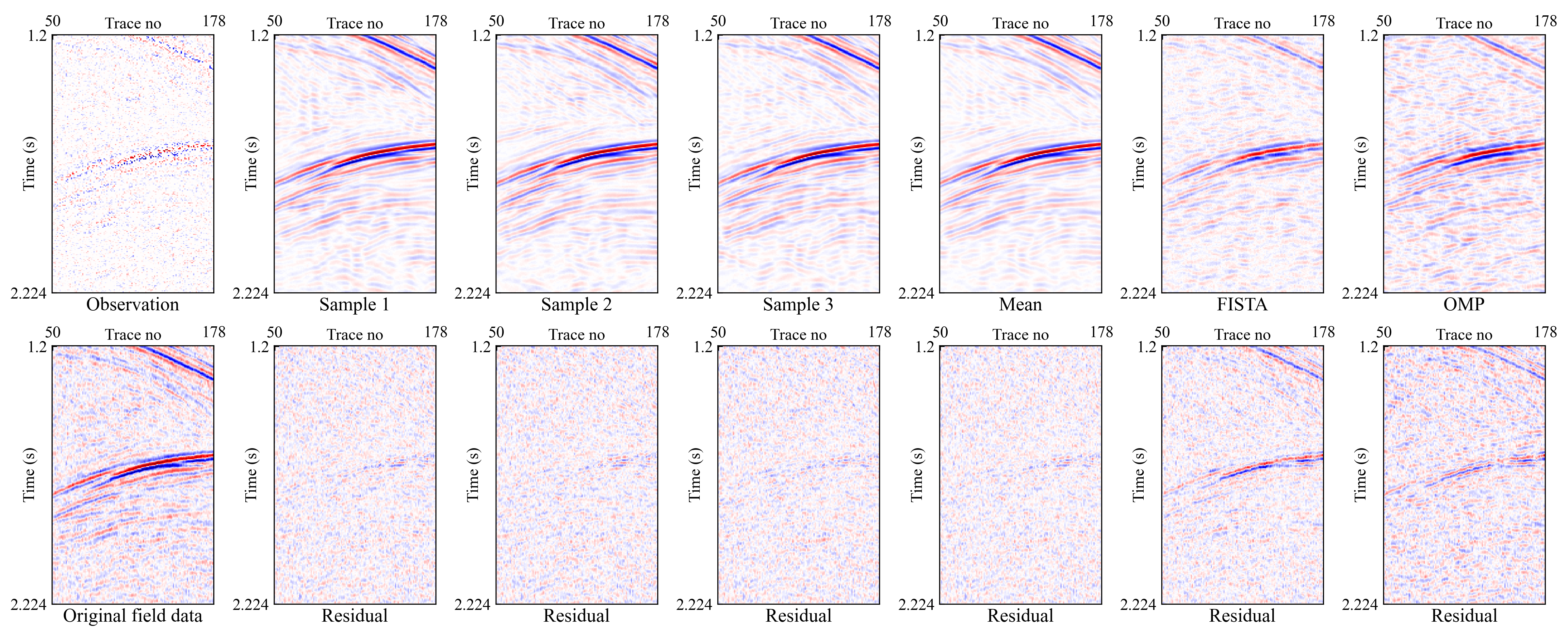}
	\caption{Unsupervised posterior sampling results, taking the compressed Sensing task as an example}
	\label{fig:XJ_cs}
	%	\vspace{-2mm}
\end{figure}

Figure~\ref{fig:XJ_cs} shows the seismic data reconstruction results under a compression ratio of 25\%. 
We employ Orthogonal Matching Pursuit (OMP)\cite{tropp2007signal} and the Fast Iterative Shrinkage-Thresholding Algorithm (FISTA)\cite{beck2009fast} as baseline compressed sensing (CS) methods. 
For OMP, the sparsifying dictionary is constructed using the Discrete Cosine Transform (DCT)~\cite{ahmed2006discrete}, and the sensing matrix $\mathbf{P}$ is a normalized random Gaussian matrix. 
Given the measurement $\mathbf{y} = \mathbf{P}\mathbf{x}$, the sparse coefficients $\boldsymbol{\alpha}$ are recovered by solving
$
\min_{\boldsymbol{\alpha}} \ \|\mathbf{y} - \mathbf{P}\mathbf{D}\boldsymbol{\alpha}\|_2^2 
\quad \text{s.t.} \quad \|\boldsymbol{\alpha}\|_0 \leq 300,
$
where $\mathbf{D}$ denotes the DCT dictionary. 
For FISTA, we implement a CS reconstruction baseline using a DCT sparsity prior and an $\ell_1$-regularized formulation:
$
\min_{\mathbf{s}} \ \frac{1}{2}\left\|\mathbf{P}\,\mathrm{IDCT2}(\mathbf{s}) - \mathbf{y}\right\|_2^2
+ \lambda \|\mathbf{s}\|_1,
$
where $\mathbf{s} = \mathrm{DCT2}(\mathbf{x})$ and $\mathbf{P}$ corresponds to a 25\% random sampling operator,  $\mathrm{IDCT2}(\cdot)$ denotes the inverse discrete cosine transform (IDCT). 
The reconstruction is performed using FISTA with a Lipschitz constant $L = 1.0$, a regularization weight $\lambda = 0.05$, and a maximum of 800 iterations. 

\subsubsection{Deconvolution}
\begin{figure}[!htb]
	%	\vspace{-2mm}
	\centering
	\includegraphics[width=\textwidth]{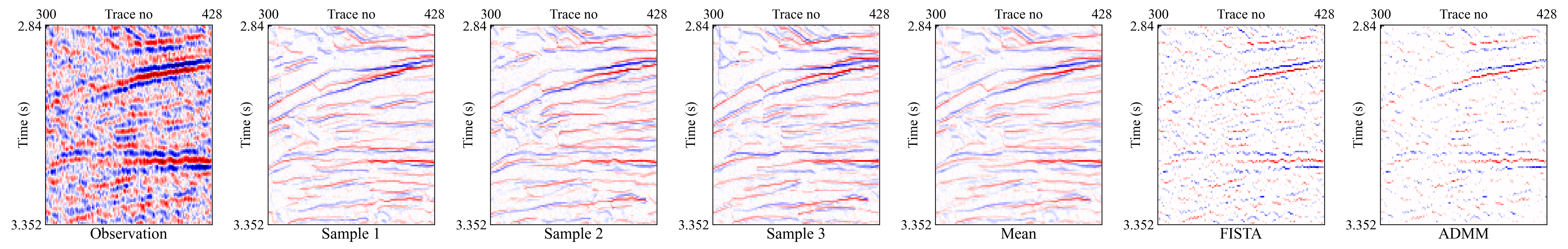}
	\caption{Unsupervised posterior sampling results, taking the deconvolution task as an example}
	\label{fig:XJ_dec}
	%	\vspace{-2mm}
\end{figure}

Figure~\ref{fig:XJ_dec} shows the deconvolution results of different methods on field seismic data. 
We adopt the FISTA\cite{beck2009fast} and the Alternating Direction Method of Multipliers (ADMM) as baseline approaches. 
The dominant frequency of the data is estimated to be $22~\mathrm{Hz}$, and a Ricker wavelet ($f_0=22~\mathrm{Hz}$, $\Delta t=4~\mathrm{ms}$) is used as the seismic wavelet. 
Each trace is modeled as the convolution of the wavelet with sparse reflectivity:
$
\mathbf{y} = \mathbf{G}\mathbf{x} + \mathbf{n},
$
where $\mathbf{G}$ denotes the Toeplitz convolution operator and $\mathbf{n}$ represents additive noise. 
Both FISTA and ADMM aim to recover the reflectivity $\mathbf{x}$ by solving the $\ell_1$-regularized least-squares problem:
$
\min_{\mathbf{x}} \frac{1}{2}\|\mathbf{G}\mathbf{x}-\mathbf{y}\|_2^2 + \lambda\|\mathbf{x}\|_1.
$
For PSF, We take predicted noise level $median(\boldsymbol{\sigma}_\mathbf{y})$=0.0657 as the value of the input noise level ${\sigma_\mathbf{y}}$.
For FISTA, the regularization parameter is set to $\lambda = 0.2$, and the algorithm is iterated for $500$ steps using the \texttt{PyLops} framework. 
For ADMM, the problem is solved in the frequency domain with symmetric zero-padding to suppress boundary artifacts, using a regularization weight $\lambda = 0.5$, and $200$ iterations. 
Both methods produce sparse reflectivity profiles that effectively enhance the temporal resolution of the recovered sections, while the deconvolved results from PSF exhibit richer fine-scale details and more continuous structures, with the stochastic samples reflecting the uncertainty of posterior sampling.

\subsection{Posterior sampling interacting with estimated noise level}

In Algorithm \ref{alg:general}, there is a hyperparameter $\sigma_\mathbf{y}$ representing the noise level to be estimated. In our method, its default value is set to the median of the pixelwise noise levels  $\boldsymbol{\sigma}_\mathbf{y}$ estimated according to Eq.\ref{Eq:sigma_map}. SGMPS-RNA~\cite{meng2025posterior} introduced the concept of interactive posterior sampling based on estimated noise level, where different settings of $\sigma_\mathbf{y}$ correspond to three modes of noise suppression: mild, moderate, and strong.

\begin{figure}[!htb]
	%	\vspace{-2mm}
	\centering
	\includegraphics[width=\textwidth]{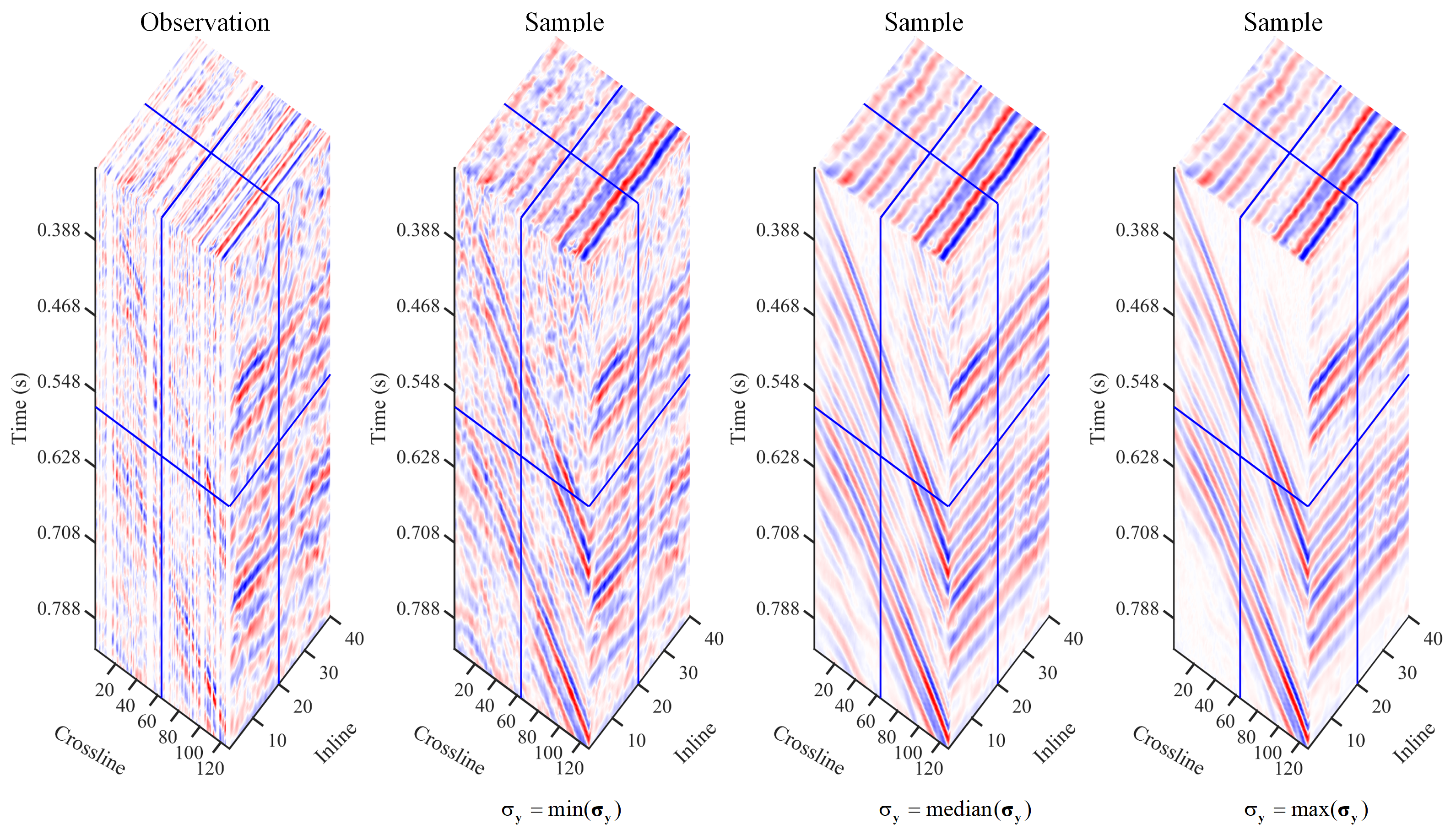}
	\caption{Posterior sampling interacting with estimated noise level}
	\label{fig:interactive sampling}
	%	\vspace{-2mm}
\end{figure}

Analogously, in PSF, $\sigma_\mathbf{y}$ is not merely a hyperparameter or a burden, but a control knob that modulates the strength of noise suppression during posterior sampling for different inverse problems. Figure~\ref{fig:interactive sampling} illustrates a result of applying PSF profile-by-profile to process a 3D field dataset with noise and missing seismic traces. When $\sigma_\mathbf{y}$ is set to a small value of $\boldsymbol{\sigma}_\mathbf{y}$, PSF mildly suppresses noise while recovering fine details of the missing seismic data. Conversely, when $\sigma_\mathbf{y}$ is set to the maximum value of $\boldsymbol{\sigma}_\mathbf{y}$, PSF strongly suppresses noise, yielding smoother reconstructions of the missing traces. Setting $\sigma_\mathbf{y}$ to the median of $\boldsymbol{\sigma}_\mathbf{y}$ provides a balanced trade-off between noise suppression and preservation of useful signal details. Therefore, similar to different stochastic solutions, $\sigma_\mathbf{y}$ offers users the flexibility to select the most appropriate reconstruction according to practical requirements.

\section{Discussion}
\begin{figure}[!htb]
	%	\vspace{-2mm}
	\centering
	\includegraphics[width=\textwidth]{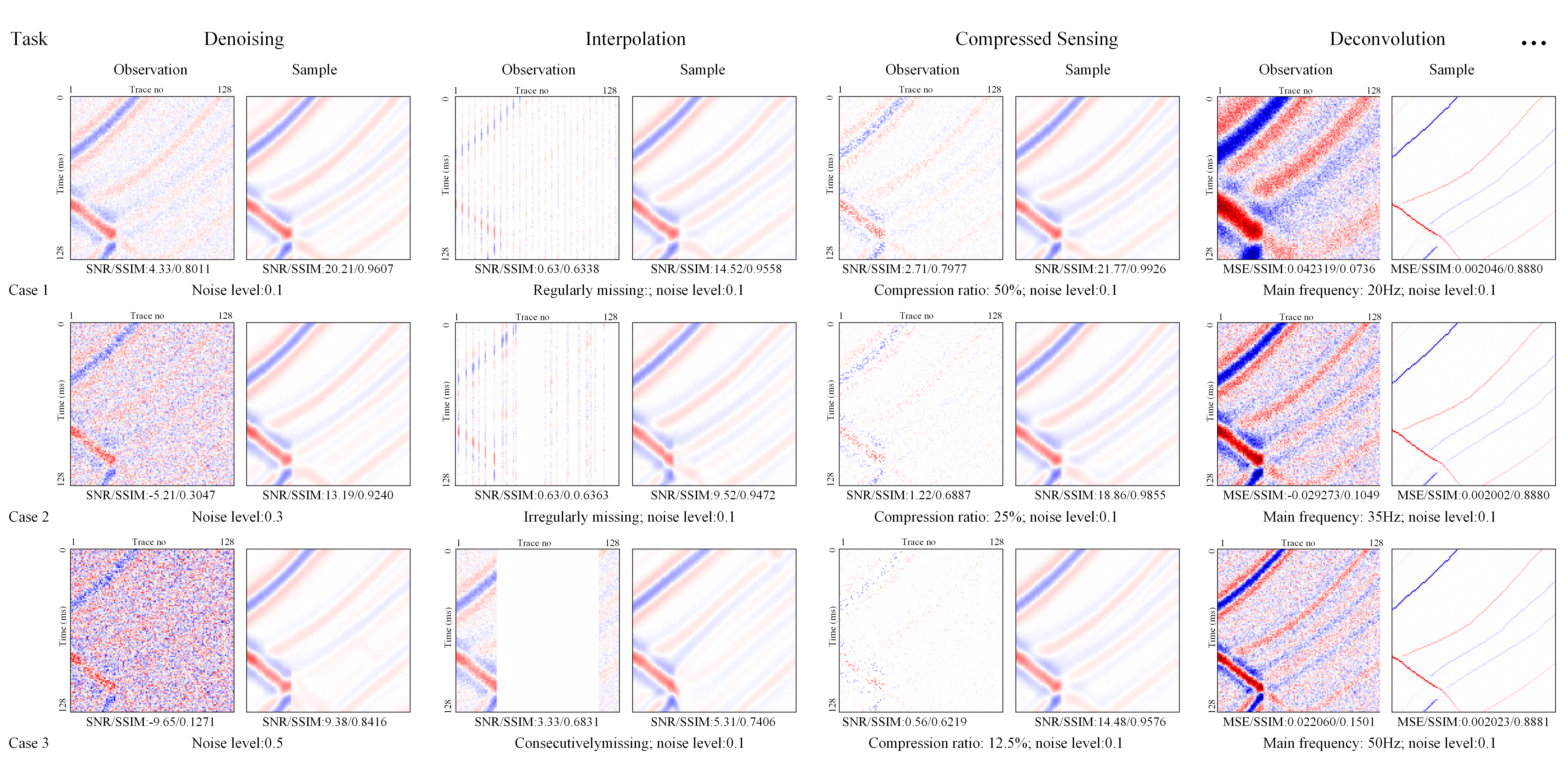}
	\caption{Posterior sampling under different signal-to-noise ratios and different degradation operators, the ground truch for seismic data and reflection coefficients is shown in Figure \ref{fig:gt_syn}.}
	\label{fig:ps_case}
	%	\vspace{-2mm}
\end{figure}

PSF is a flexible, unsupervised posterior sampling framework. As shown in Figure \ref{fig:ps_case}, we demonstrate seismic data recovery under varying signal-to-noise ratios, different missing-trace patterns, diverse compression ratios, and convolutions with wavelets of different dominant frequencies. By automatically estimating $\sigma_y$ and adjusting the operator $\mathbf{G}$, PSF can handle seismic data with unknown noise levels and unknown degradation processes.  For more complex nonlinear inverse problems, $\mathbf{G}$ can first be linearized before applying PSF for reconstruction. Since PSF requires knowledge of the forward operator $\mathbf{G}$, new posterior sampling algorithms are needed when $\mathbf{G}$ is unknown, such as in blind deconvolution tasks with unknown wavelets. Moreover, because PSF posterior sampling involves hundreds of function evaluations, developing accelerated sampling strategies will be essential. These directions represent promising avenues for future research.

\begin{figure}[!htb]
	%	\vspace{-2mm}
	\centering
	\includegraphics[width=0.4\textwidth]{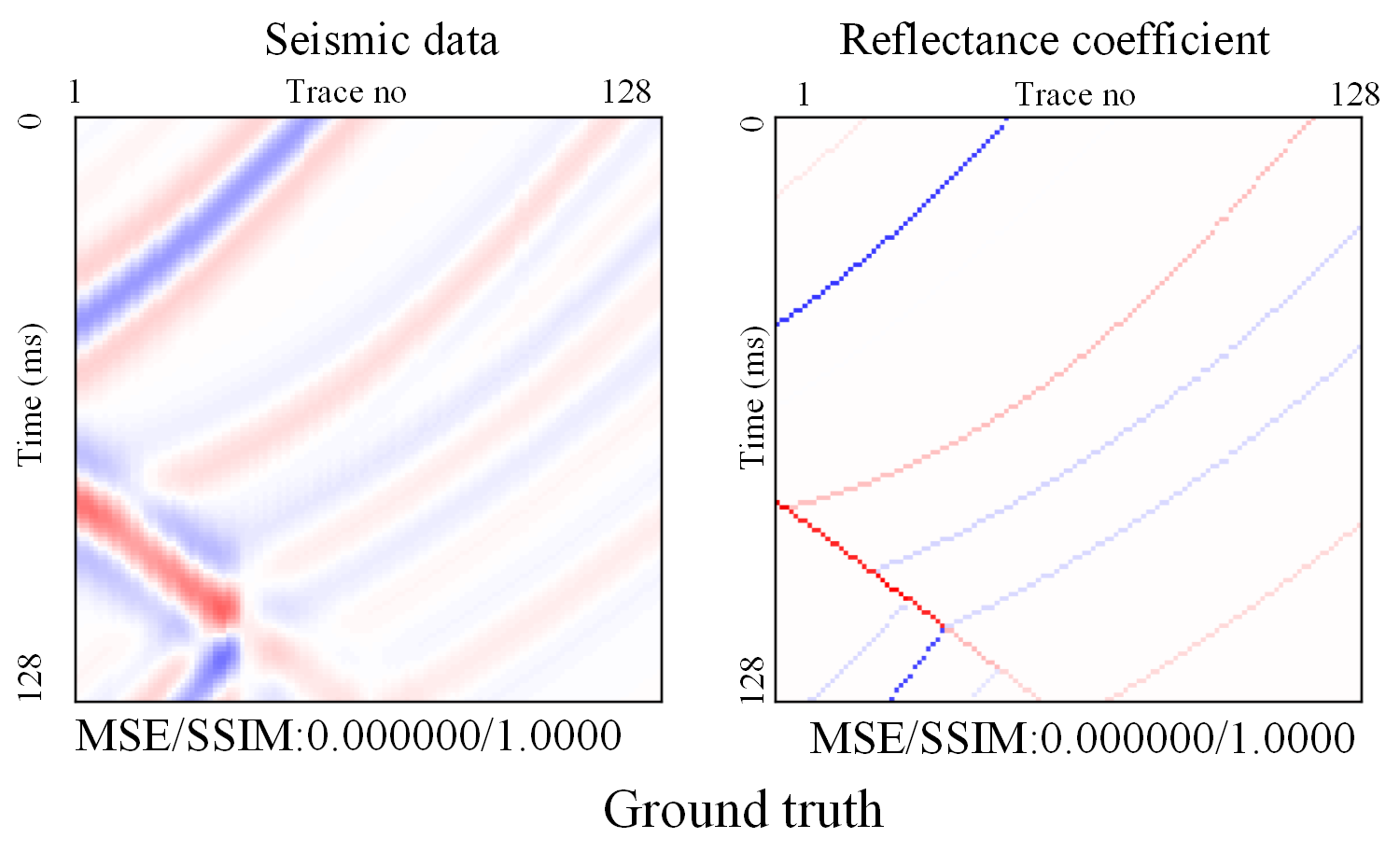}
	\caption{the ground truch for seismic data and reflection coefficients in Figure \ref{fig:ps_case}}
	\label{fig:gt_syn}
	%	\vspace{-2mm}
\end{figure}

%\begin{figure}[!htb]
%	%	\vspace{-2mm}
%	\centering
%	\includegraphics[width=0.8\textwidth]{images/ps_case_more.png}
%	\caption{Posterior sampling examples for more tasks}
%	\label{fig:ps_case_more}
%	%	\vspace{-2mm}
%\end{figure}

\section{Conclusion} 
This paper proposes an unsupervised PSF for seismic data restoration.
PSF leverages the generative prior of a pre-trained unconditional SGMs to derive a generalized conditional score function associated with the forward operator of different inverse problems.
By incorporating an automatic noise-level estimation strategy, PSF enables adaptive control of noise suppression strength during posterior sampling.
Consequently, PSF achieves posterior sampling for various inverse problems under arbitrary  degradation processes and signal-to-noise ratios without any model retraining.
Posterior sampling experiments demonstrate that PSF effectively leverages the rich seismic-aware generative priors inherently encoded in trained unconditional SGMs and exhibits strong out-of-distribution generalization across various types of field seismic data. Furthermore, experiments on denoising, interpolation, compressed sensing, deconvolution, and other inverse problems confirm the superior adaptability and robustness of PSF across diverse tasks.

\bibliographystyle{unsrt}  
\bibliography{ref}

\end{document}